\documentclass[fleqn,usenatbib]{mnras}

\usepackage[T1]{fontenc}

\DeclareRobustCommand{\VAN}[3]{#2}
\let\VANthebibliography\thebibliography
\def\thebibliography{\DeclareRobustCommand{\VAN}[3]{##3}\VANthebibliography}

\usepackage{graphicx}	
\usepackage{amsmath}	
\usepackage{amssymb}	
\usepackage{xcolor}     

\usepackage{comment}
\usepackage{newtxtext,newtxmath}

\usepackage{lineno}

\newcommand{\fnl}{f_{\rm NL}}
\newcommand{\keq}{k_{\rm eq}}
\newcommand{\kmin}{k_{\rm min}}
\newcommand{\rmax}{r_{\rm max}}
\newcommand{\shi}{{\rm shi}}
\newcommand{\CHI}{{\rm chi}}
\newcommand{\goliat}{\textsc{goliat-png} }

\def\p{{\bf p}}

\def\r{{\bf r}}
\def\x{{\bf x}}

\defcitealias{2018Y1}{DES Collaboration 2018a}
\defcitealias{descollaboration2021darkY3}{DES Collaboration 2021a}
\defcitealias{2018BAOY1}{DES Collaboration 2018b}
\defcitealias{descollaboration2021dark}{DES Collaboration 2022b}
\defcitealias{2021DR2}{DES Collaboration 2021}

\usepackage{eso-pic}

\AddToShipoutPictureBG*{%
  \AtPageUpperLeft{%
    \hspace{0.75\paperwidth}%
    \raisebox{-4.6\baselineskip}{%
      \makebox[0pt][l]{\textnormal{DES-2021-0676}}
}}}%

\AddToShipoutPictureBG*{%
  \AtPageUpperLeft{%
    \hspace{0.75\paperwidth}%
    \raisebox{-5.6\baselineskip}{%
      \makebox[0pt][l]{\textnormal{FERMILAB-PUB-22-655-V}}
}}}%

\AddToShipoutPictureBG*{%
  \AtPageUpperLeft{%
    \hspace{0.75\paperwidth}%
    \raisebox{-6.6\baselineskip}{%
      \makebox[0pt][l]{\textnormal{IFT-UAM/CSIC-22-107}}
}}}%

\title[PNG with ACF]{Primordial non-Gaussianity with Angular correlation function:\\ Integral constraint and validation for DES}

\author[W. Riquelme et al.]{
\parbox{\textwidth}{
Walter Riquelme,$^{1, 2}$\thanks{E-mail: walter.riquelme@uam.es}
Santiago Avila,$^{1,2}$\thanks{E-mail: santiago.avila@uam.es}
Juan Garc\'ia-Bellido,$^{1,2}$\thanks{E-mail: juan.garciabellido@uam.es}
Anna Porredon,$^{3, 4, 5}$
Ismael Ferrero,$^{6}$
Kwan Chuen Chan,$^{7, 8}$
Rogerio Rosenfeld,$^{9}$
Hugo Camacho,$^{10, 11}$
Adrian G. Adame,$^{1, 2}$
Aurelio Carnero Rosell,$^{11, 12}$
Martin Crocce,$^{13, 14}$
Juan De Vicente,$^{15}$
Tim Eifler,$^{16, 17}$
Jack Elvin-Poole,$^{3, 4}$
Xiao Fang,$^{18, 19}$
Elisabeth Krause,$^{19}$
Martin Rodriguez Monroy,$^{15}$
Ashley J. Ross,$^{3}$
Eusebio Sanchez,$^{15}$
Ignacio Sevilla$^{15}$
}
\vspace{0.4cm}
\\
$^{1}$ Instituto de F\'isica Teorica UAM-CSIC, c/ Nicolás Cabrera 13-15, 28049 Madrid, Spain \\
$^{2}$ Departamento de F\'isica Te\'orica,  Universidad Aut\'onoma de Madrid, 28049 Madrid, Spain \\
$^{3}$ Center for Cosmology and Astro-Particle Physics, The Ohio State University, Columbus, OH 43210, USA \\
$^{4}$ Department of Physics, The Ohio State University, Columbus, OH 43210, USA \\
$^{5}$ Institute for Astronomy, University of Edinburgh, Edinburgh EH9 3HJ, UK \\
$^{6}$ Institute of Theoretical Astrophysics, University of Oslo. P.O. Box 1029 Blindern, NO-0315 Oslo, Norway \\
$^{7}$ School of Physics and Astronomy, Sun Yat-sen University, 2 Daxue Road, Tangjia, Zhuhai, 519082, China \\
$^{8}$ CSST Science Center for the Guangdong-Hongkong-Macau Greater Bay Area, SYSU, China \\
$^{9}$ ICTP South American Institute for Fundamental Research, Instituto de F\'{\i}sica Te\'orica, Universidade Estadual Paulista, S\~ao Paulo, Brazil \\
$^{10}$ Instituto de F\'{i}sica Te\'orica, Universidade Estadual Paulista, S\~ao Paulo, Brazil \\
$^{11}$ Laborat\'orio Interinstitucional de e-Astronomia - LIneA, Rua Gal. Jos\'e Cristino 77, Rio de Janeiro, RJ - 20921-400, Brazil\\
$^{12}$ Instituto de Astrofisica de Canarias, E-38205 La Laguna, Tenerife, Spain \\
$^{13}$ Institut d'Estudis Espacials de Catalunya (IEEC), 08034 Barcelona, Spain \\
$^{14}$ Institute of Space Sciences (ICE, CSIC),  Campus UAB, Carrer de Can Magrans, s/n,  08193 Barcelona, Spain \\
$^{15}$ Centro de Investigaciones Energ\'eticas, Medioambientales y Tecnol\'ogicas (CIEMAT), Madrid, Spain \\
$^{16}$ Department of Astronomy/Steward Observatory, University of Arizona, 933 North Cherry Avenue, Tucson, AZ 85721-0065, USA \\
$^{17}$ Jet Propulsion Laboratory, California Institute of Technology, 4800 Oak Grove Dr., Pasadena, CA 91109, USA \\
$^{18}$ Department of Astronomy, University of California, Berkeley,  501 Campbell Hall, Berkeley, CA 94720, USA \\
$^{19}$ Department of Astronomy/Steward Observatory, University of Arizona, 933 North Cherry Avenue, Tucson, AZ 85721-0065, USA \\
}

\date{Accepted XXX. Received YYY; in original form ZZZ}

\pubyear{2022}

\begin{document}
\label{firstpage}
\pagerange{\pageref{firstpage}--\pageref{lastpage}}
\maketitle


\begin{abstract}
Local primordial non-Gaussianity (PNG) is a promising observable of the underlying physics of inflation, characterised by $\fnl^{\rm loc}$. We present the methodology to measure $\fnl^{\rm loc}$ from the Dark Energy Survey (DES) data using the 2-point angular correlation function (ACF) with scale-dependent bias.
One of the focuses of the work is the integral constraint. This condition appears when estimating the mean number density of galaxies from the data and is key in obtaining unbiased $\fnl^{\rm loc}$ constraints.
The methods are analysed for two types of simulations: $\sim 246$ \goliat N-body small area simulations with $\fnl$ equal to -100 and 100, and 1952 Gaussian ICE-COLA mocks with $\fnl=0$ that follow the DES angular and redshift distribution.
We use the ensemble of \goliat mocks to show the importance of the integral constraint when measuring PNG, where we recover the fiducial values of $\fnl$ within the $1\sigma$ when including the integral constraint. In contrast, we found a bias of $\Delta \fnl\sim 100$ when not including it. For a DES-like scenario, we forecast a bias of $\Delta \fnl \sim 23$, equivalent to $1.8\sigma$, when not using the IC for a fiducial value of $\fnl=100$.
We use the ICE-COLA mocks to validate our analysis in a realistic DES-like setup finding it robust to different analysis choices: best-fit estimator, the effect of IC, BAO damping, covariance, and scale choices.
We forecast a measurement of $\fnl$ within $\sigma(\fnl)=31$ when using the DES-Y3 BAO sample, with the ACF in the $1\ {\rm deg}<\theta<20\ {\rm deg}$ range.
\end{abstract}


\newpage

\section{Introduction}
Cosmic inflation predicts that the primordial seeds, encoded in the initial gravitational potential of the Universe, are described by close to Gaussian random fields, for which all the statistical information is contained in the two-point correlation function. We can parametrise deviations from Gaussianity by using a parameter denoted by $\fnl$, which represents the amount of primordial non-Gaussianity encoded in the three-point correlation of the fields.
Primordial non-Gaussianity (PNG) is claimed to be a smoking gun to differentiate among the vast collection of inflationary models. In particular, primordial non-Gaussianity of the local type, parametrised by $\fnl^{\rm loc}$, can distinguish between canonical single-field and non-vanilla scenarios, such as multi-field inflation \citep{2013pajer, 2010multi}.

The primordial seeds affect the formation of structures at different epochs in cosmic history, implying that signals of PNG could appear in different cosmological probes. An example is the constraints of PNG coming from the cosmic microwave background (CMB) temperature bispectrum. The latest Planck results present the tightest constraints for local PNG with $\fnl^{\rm loc}=-0.9\pm5.1$ \citep{2020A&A...641A...9P}, but since Planck reached its cosmic variance limit, another way to improve this constraint is desirable.

Similar to how PNG affects the temperature fluctuations in the CMB, the non-Gaussian initial perturbations can also affect the distribution of dark matter overdensities, which in turn affects the distribution of biased tracers of dark matter (e.g., galaxies, quasars). This implies that PNG could also be constrained using the bispectrum of such tracers, as has been studied in \citet{2009KomatsuLSS, 2014TasinatoBS, 2021Azadeh}. 

Given the complexity of modelling the bispectrum, dominated by late non-Gaussianities induced by non-linear evolution\footnote{It is worth mentioning that besides these difficulties, recent work using the EFT of LSS for the bispectrum has proven to be helpful when constraining local PNG from eBOSS data \citep{2022PhRvD.106d3506C}.} and other difficulties such as non-linear bias, redshift space distortions, and the window function of the survey \citep{2017MNRAS.465.1757G, 2019MNRAS.484..364S}, a different method to look for primordial non-Gaussianity using late-time objects is desired. Another effect of PNG is on the halo formation mechanism. Local primordial non-Gaussianity induces a scale dependence on the linear bias between galaxies and the underlying dark matter over-densities. The scale dependence in the bias creates a characteristic signal in the two-point correlation at very large scales, which can be constrained using different large-scale structure (LSS) biased tracers. \citep{2008Dalal, 2008Slosar, 2008Matarrese}. Some studies show that PNG can also be constrained using galaxies with zero linear bias in low-density environments \citep{2018Castorina}, or even negative biased traces, such as voids \citep{2019PhRvD..99l1304C}.

Measurements of cosmological parameters using two-point correlation functions have been done multiple times because they are easy to model and have a large signal-to-noise ratio. This makes the scale-dependent bias in the two-point correlation the more robust method to constrain PNG.
Previous measurements of PNG using the scale-dependent bias have been presented in \citet{2008Slosar, Ross2013, PhysRevD.89.023511, 2015Ho, PhysRevLett.113.221301, 2019Castorina, mueller2021clustering}.

One noticeable trend is that most of the current constraints come from spectroscopic surveys.
It has been shown in \citet{PhysRevD.95.123513} that imaging surveys with high volumes could overcome redshift uncertainties and had the potential of breaking the $\sigma(\fnl^{\rm loc})\sim1$ barrier.
Hence, upcoming photometric data from the Legacy Survey of Space and Time (LSST) in the Vera Rubin Observatory \footnote{\url{https://www.lsst.org/}} \citep{lsstsciencecollaboration2009lsst} is a promising source to break current bounds.

This work is a first step to measure PNG with existing data from the Dark Energy Survey (DES)\footnote{\url{https://www.darkenergysurvey.org/}}  \citepalias{2021DR2}, which represents the state of the art in photometric surveys. Currently, the DES has surveyed over $\sim 388$ million galaxies in $\sim 5000\ {\rm deg}^{2}$ and presents an opportunity to put the tightest constrains from photometric surveys (as will see in this work).

DES has successfully probed the nature of dark energy using different cosmological probes (\citetalias{2018Y1, descollaboration2021darkY3}; \citet{porredon2021dark, rodriguezmonroy2021dark}). One of them is the study of clustering of galaxies for the measurement of the Baryon Acoustic Oscillation (BAO) scale \citepalias{2018BAOY1, descollaboration2021dark} using galaxy data.
The BAO scale measurement suggests that we could also use clustering of galaxies at large scales for measuring PNG within DES.

This work presents the starting point in this direction by describing the methods to constrain the $\fnl^{\rm loc}$ parameter using DES simulations. We use the angular correlation function (ACF) as a summary statistic for the galaxy distribution and show the effect that primordial non-Gaussianities have on the angular clustering of galaxies via the scale-dependent bias.

One of the main focuses of the work is on the integral constraint (IC) \citep{1977ApJ...217..385G, 10.1093/mnras/253.2.307, 2014icBeutler, Ross2013, 2019ic}. The integral constraint corrects the modelled correlation function by adding a constant, which comes from imposing that its integral over the whole survey volume needs to vanish. This correction is found to be key to obtaining unbiased PNG measurements.

The integral constraint was not relevant in the previous DES non-PNG clustering analysis for two main reasons: First, its effect becomes relevant at very large scales.
Secondly, for the case of BAO measurements, its template includes marginalisation over nuisance parameters, one of them being a constant shift in the amplitude of the ACF. This shift mimics the integral constraint correction, implying that any effect from it has already been marginalised.

In this paper, we use the angular correlation function with PNG, and the integral constraint, as a theoretical template to measure the value of $\fnl^{\rm loc}$ from simulated galaxy catalogues. The measurement is based on Bayesian parameter inference using MCMC (Markov chain Monte Carlo) sampling of a Gaussian likelihood function. The methods are analysed for two kinds of simulations. First, we introduce the \goliat mocks \citep{2023MNRAS.519.3706A}, a set of 246 N-Body simulations that have non-Gaussian initial conditions. We use these simulations to remark on the importance of the integral constraint when measuring $\fnl^{\rm loc}$.
Second, we use 1952 ICE-COLA mocks \citep{DES:2021fie} that follow the DES angular and redshift distribution of the Y3 BAO galaxy sample \citep{rosell2021dark} to validate the pipeline. We show that it is robust against different analysis choices, such as covariance modelling, $\fnl$ estimator, and scale cuts. Finally, we forecast a measurement of the accuracy of $\fnl^{\rm loc}$ when using the DES Y3 BAO sample data.


This paper is organised as follows. The steps to model angular correlation function with scale-dependent bias are presented in Section \ref{sec:theory}. In Section \ref{sec:ic}, we derive the integral constraint and show its importance when dealing with local PNG. In Section \ref{sec:sims}, we describe the simulations that we will use to test and optimise the methods. Section \ref{sec:tools} presents the tools needed to extract the $\fnl$ parameter. In Section \ref{sec:test_goliat}, we test the pipeline against the \goliat simulations and show how the integral constraint is needed to obtain unbiased values of $\fnl$. Once the methods are tested over non-Gaussian simulations, we validate the pipeline using ICE-COLA simulations in Section \ref{sec:opt_cola}. 

\section{Theory}\label{sec:theory}

In this section, we describe the impact of PNG on the two-point statistics of biased tracers. First, we describe how non-Gaussian initial conditions modify the bias relation, introducing the scale-dependent bias. After, we show the effect that it has on the power spectrum. Finally, we focus on the angular correlation function and show how it is affected by local Primordial non-Gaussianity.

\subsection{Gaussian galaxy bias}\label{ssec:scale_bias}

The spatial distribution of matter is set by the initial conditions coming from cosmic inflation, which predicts a nearly scale-invariant power spectrum and a close to Gaussian distribution for the primordial gravitational fields.
During the matter domination era, dark matter collapsed due to these gravitational potentials generating halos which, as the Universe evolves, will serve as the backbones for the creation of large-scale structures.

We will focus our analysis on angular separations of galaxies larger than 1 degree. This choice is customary for the BAO analysis because such scales are within the linear regime of perturbation theory, simplifying the theoretical modelling \citep{descollaboration2021dark}. In this regime, galaxies follow the trace of the dark matter overdensities by the linear relation,
\begin{equation}\label{eq:b}
    \delta_{\rm g}(\x) = b\ \delta_{\rm m}(\x),
\end{equation}
where $b$ is a parameter called galaxy bias, which is found to be constant at large scales under the standard Gaussian initial conditions.

In the non-linear regime, non-linear effects also generate a scale-dependent bias, which affects only small scales. We will ignore such effects throughout this work and refer the reader to \citet{Desjacques_2018} for an intensive review on the scale dependence of the galaxy bias and other related effects.

The statistical distribution of dark matter overdensities is well described by the matter power spectrum $P_{\rm m}(k)$, which depends on the primordial power spectrum, coming from inflation, and the transfer function $T(k)$, which describes its evolution throughout cosmic history. Due to Equation \ref{eq:b}, the biased relation between galaxies and dark matter also appears in the galaxy power spectrum, as follows,
\begin{equation}
    P_{\rm g}(k) = b^{2}P_{\rm m}(k).
\end{equation}

As we will see in the following section, the linear relation between galaxies and dark matter will change when dealing with non-Gaussian initial conditions.

\subsection{PNG via scale-dependent bias}\label{ssec:scale_bias}
Deviations from Gaussianity in the initial conditions, coming from inflation, is an active area of research due to the potential of unveiling the nature of the primordial fields. In particular, we focus on PNG of the local type \citep{2001Komatsu},
\begin{equation} \label{eq:fnl}
        \Phi_{\rm NG}(\x) = \phi_{\rm G}(\x) + \fnl^{\rm loc}(\phi_{\rm G}^{2}(\x)-\langle\phi_{\rm G}^{2}\rangle),
\end{equation}
where $\Phi_{\rm NG}(\x)$ is the non-Gaussian Newtonian potential and $\phi_{G}(\x)$ is the Gaussian potential. Under this approximation, $\fnl^{\rm loc}$ is a constant that parametrises deviations from Gaussian initial conditions. Throughout this work, we will focus on local PNG; hence, from here on, we will drop the superscript 'loc' for simplicity.

\citet{2008Dalal} and \citet{2008Slosar} showed that PNG, parametrised as Eq.(\ref{eq:fnl}), would change the way dark matter collapses into halos, subsequently affecting galaxy formation. In the presence of local PNG, the long wavelength modes of the primordial gravitational potential couple with the smaller modes, responsible for the local amplitude of matter fluctuations, producing a modulation in the local number density of halos. The change in the local number density will add an extra contribution to the galaxy bias, which depends on the scale. We can write the scale-dependent bias due to local PNG as follows,
\begin{equation} \label{eq:scale_dep}
    b(k) = b + \fnl \alpha(k, z)\frac{\partial \ln n}{\partial \ln \sigma_{8}},
\end{equation}
where $b$ is the constant linear bias and $\delta_{c}=1.686$ is the critical value of collapse for halo formation in an Einstein-de Sitter universe \citep{1984ApJ...281....1F}. 
Also,
\begin{equation}\label{eq:alpha}
    \alpha(z, k)=\frac{3\Omega_{m}}{2D(z)}\frac{H_{0}^{2}}{c^{2}}\frac{g(0)}{g(z_{\rm rad})}\frac{1}{k^{2}T(k)},
\end{equation}
where $H_{0}$ is the Hubble factor today \footnote{If one uses k in units of $h \text{Mpc}^{-1}$, then $H_{0}=100h[{\rm Mpc}^{-1}{\rm km}\ {\rm s}^{-1}]$ with $h=0.7$ }, $c$ the light speed and $\Omega_m$ the matter density today.
In addition, $T(k)$ is the linear transfer function, and $D(z)$ is the linear growth factor, both normalised to 1 at $k=0$ and $z=0$, respectively. The factor $\frac{g(0)}{g(z_{\rm rad})}$, with $g(z)=(1+z)D(z)$, arises because $D(z)$ is normalised to unity and can be omitted if normalised to the scale factor during the matter-dominated era \citep{2018Mueller}. Its value is shown to be $\frac{g(0)}{g(z_{\rm rad})}\simeq1.3$\footnote{This value is slightly cosmology dependent. When comparing against the ICE-COLA mocks, we will consider it as $1.3$ since we do not expect that it affects the constraints if we plan to recover $\fnl=0$. On the other side, for the non-Gaussian \goliat simulations, it was shown to be $1.316$ for the fiducial cosmology of the simulations.}.

One particularity of this scale-dependent bias is its $1/k^{2}$ dependence, implying that primordial non-Gaussianity affects the distribution of galaxies only at very large scales. Throughout this work, we will refer to scale-dependent bias as the one produced due to primordial non-Gaussianity.

It has been shown in \citet{2008Slosar} that,
\begin{equation}
    \frac{\partial \ln n}{\partial \ln \sigma_{8}} = 2\delta_{c}(b-p)
\end{equation}
where the parameter $p$ was introduced to show deviations from the original model of \citet{2008Dalal} to take into account different tracers. We refer the reader to \citet{2020abarreira} for an analysis of the impact of the parameter $p$ and other assumptions on the non-Gaussian bias. For the case of ICE-COLA mocks, we will fix $p=1$, which is customary in many analyses and is considered the prediction for a mass-selected galaxy/halo sample. Finally, the scale-dependent bias we will use in this work can be written as follows, 
\begin{equation} \label{eq:scale_dep}
    b(k) = b + 2(b-p)\fnl \alpha(k, z) \delta_{c}.
\end{equation}

\begin{figure}
	\includegraphics[width=\columnwidth]{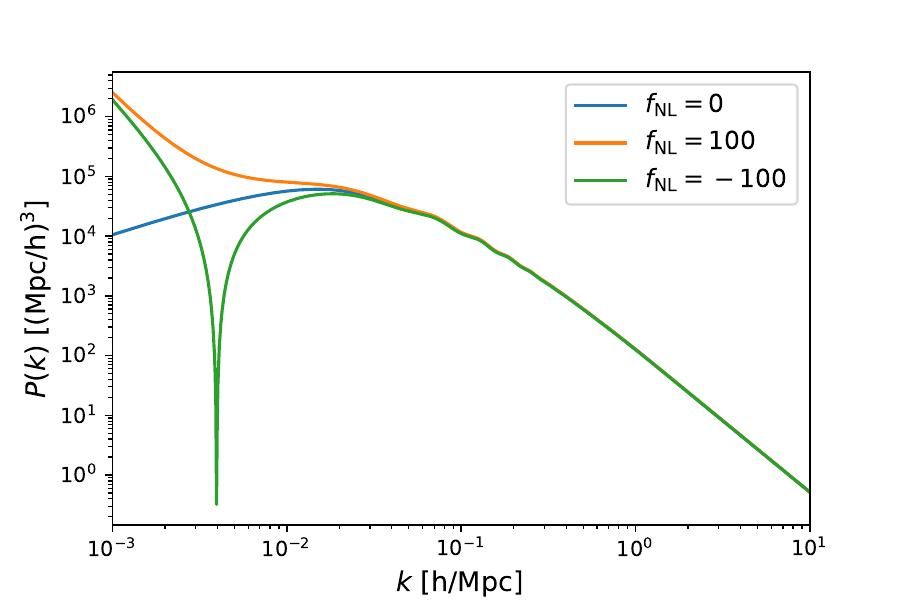}
    \caption{Theoretical linear galaxy power spectrum with scale-dependent bias for $\fnl=0$ (blue line), $\fnl=100$ (orange line) and $\fnl=-100$ (green line). The power spectrum is computed using the fiducial cosmological parameters of the \goliat simulations described in Table \ref{tab:param}.}
    \label{fig:pk_bias}
\end{figure}

As an example of the effect of the scale-dependent bias, in Figure \ref{fig:pk_bias}, we compute the linear matter power spectrum from CAMB\footnote{\url{https://camb.readthedocs.io}} \citep{Lewis:1999bs, Howlett:2012mh}  and apply a scale-dependent bias as given in Eq.(\ref{eq:scale_dep}) to show the galaxy power spectrum for different values of $\fnl$. The power spectrum is computed using the cosmological parameters from the ICE-COLA simulation presented in subsection \ref{ssec:cola}.

Since the scale-dependent bias is squared in the galaxy power spectrum, we will have contributions with different dependence on $\fnl$. This dependence can be seen as follows,
\begin{equation}\label{eq:bias_fnl}
    b(k)^{2} \propto b^{2} + A\ b\frac{\fnl}{k^{2}} + B\frac{\fnl^{2}}{k^{4}},
\end{equation}
where $A$ and $B$ are prefactors that do not depend on the scale (since $T(k)$ becomes constant at very large scales). The previous equation tells us that we have quadratic and linear terms in $\fnl$ and a term that does not depend on $\fnl$. Figure \ref{fig:pk_bias} shows how the scale-dependent bias generates an enhancement of the power spectrum at large scales for $\fnl=100$. The situation is more interesting for $\fnl=-100$, where the linear term in $\fnl$ generates a reduction in the power spectrum until a given scale, then the quadratic term overcomes, explaining the sharp feature around at $k=0.005h{\rm Mpc}^{-1}$.

\subsection{BAO-damped galaxy power spectrum}
We may need to use precise theoretical modelling to obtain an optimal measurement of $\fnl$. For this, we follow the methodology used in \citetalias{descollaboration2021dark} for the DES Y3 BAO template, based on extensions of the linear power spectrum using IR resummation methods optimised for an accurate description of the damping in the BAO peak \citep{2016Blas, 2018Ivanov}. The particularity of this method relies on a derivation of the BAO damping based on first principles, in contrast with other models where the damping is obtained from fits over simulations. In subsection \ref{ssec:lin_theory}, we will compare the impact of using the BAO-damped galaxy power spectrum versus linear theory without damping on the $\fnl$ measurement. 

The BAO-damped galaxy power spectrum is given by:
\begin{equation}\label{eq:non_lin_ps}
\begin{split}
    P(k, \mu, z)&=(b(k) + f(z)\mu^{2})^{2}\left[(P_{\rm lin}(k)-P_{\rm nw}(k))D_{\rm BAO} + P_{\rm nw}(k) \right],
\end{split}
\end{equation}
where $P_{\rm lin}(k)$ is the linear matter power spectrum. $P_{\rm nw}(k)$ is the smooth "no-wiggle" power spectrum. We refer the reader to \citetalias{descollaboration2021dark} for further details on how to compute it. The function $f(z)$ is the growth rate of structures, defined under the following approximation \citep{PhysRevD.72.043529},
\begin{equation}
    f(z)\approx\Omega_{m}(z)^{\gamma},
\end{equation}
with $\gamma=0.55$.
The parameter $\mu$ is defined as the cosine of the angle between the line of sight and wave vector $\mathbf{k}$.

In Equation \ref{eq:non_lin_ps}, $D_{\rm BAO}(z)$ is a Gaussian damping defined by:
\begin{equation}\label{eq:gaussian_damp}
\begin{split}
D_{\rm BAO}(z)&=\exp\{-k^{2}(\mu^2 \Sigma_{\parallel}^{2} + (1-\mu^{2})\Sigma_{\perp}^{2} + f(z) \mu^{2}(\mu^{2}-1)\delta\Sigma^{2}), \} 
\end{split}
\end{equation}
where $\Sigma_{\parallel}(z)=(1+f(z)\Sigma_{\perp})$. The parameters $\Sigma_{\perp}$ and $\delta\Sigma$ can be computed directly for a fixed cosmology. In the case of ICE-COLA cosmology, at $z=0$, $\Sigma_{\perp}=5.8 {\rm Mpc}/h$ and $\delta\Sigma = 3.18 {\rm Mpc}/h$ and they are scaled by the growth factor to any other redshift \citepalias{descollaboration2021dark}.

When comparing against the ICE-COLA simulations, we will include the BAO damping in the power spectrum, as presented in this subsection, since we will be using these simulations to validate the methods and improve the accuracy for $\fnl$, implying the need for a more precise theory modelling. 
When comparing against $\goliat$ simulations, we will not consider BAO damping because we use those simulations to recover higher $\fnl$ values, and we do not expect the damping to be a determinant factor in their accuracy.
We will come back to this discussion on subsection \ref{ssec:lin_theory}, where we will assess the impact of the BAO damping on the $\fnl$ measurement.
Also, notice that the scale-dependent bias described in the previous subsection is already added in Eq.(\ref{eq:non_lin_ps}), adding extra contributions to the galaxy power spectrum.

With the previously computed power spectrum, we can use a multipole expansion in Legendre polynomials of $\mu$,
\begin{equation}\label{eq:multi_ps}
P_{\ell}(k,z) \equiv  \frac{(2\ell+1)}{2}\int_{-1}^{1} \text{d}\mu P(k,\mu,z)L_{\ell}(\mu),
\end{equation}
to take into account the anisotropies caused by redshift space distortions to the line of sight. Notice that the power spectrum is computed at $z=0$ and does not include the growth factor $D(z)$ since this will be added when calculating the angular correlation function in the next section.


\subsection{Angular correlation function with PNG}\label{ssec:acf_png}
Using the previously described power spectrum, we can compute its configuration space counterpart, the two-point correlation function (2PCF), using the multipole expansion of Eq.(\ref{eq:multi_ps}),
\begin{eqnarray}\label{eq:2PCF}
\xi(r, \hat{\textbf{r}}\cdot\hat{\textbf{l}}) &=& \sum_{\ell=0,2,4} \xi_{\ell}(r)L_{\ell}(\hat{\textbf{r}}\cdot\hat{\textbf{l}}),\\
\xi_{\ell}(r) &=& \frac{i^{\ell}}{2\pi^{2}}\int_{0}^{\infty}\text{d}k\ k^{2}j_{\ell}(k r)P_{\ell}(k,\bar{z}),
\end{eqnarray}
where $r$ is the separation distance between galaxies and $j_{\ell}$ is the spherical Bessel function.
Notice that the previously computed power spectrum is evaluated at the mean redshift of the photo-z distribution, $\bar{z}$.
The correlation function is also a function of the angle between the line of sight direction $\hat{\textbf{l}}$ and the direction of the separation vector $\hat{\textbf{r}}$, given by
\begin{equation}
    \hat{\textbf{r}}\cdot\hat{\textbf{l}} = \frac{\chi(z_{2})-\chi(z_{1})}{r}\cos{\frac{\theta}{2}},    
\end{equation}
where $\chi(z)$ is the comoving distance, and $\theta$ is the angular separation between two galaxies.

It is important to remember that because of primordial non-Gaussianity, we now have a scale-dependent bias $b(k)$ that will be a part of each $P_{\ell}(k, z)$ and needs to be considered for the computation of the 2PCF.

We can compute the angular correlation function (ACF) \citep{2011Crocce, 2018Chan} as the 2-dimensional projection of the 2PCF following the galaxy photo-z distribution, $N(z)$, normalised such that its integral over redshift is equal to 1. With this, the ACF is given by:
\begin{equation} \label{eq:acf_fnl}
    w(\theta)=\int \text{d}z_{1}\int \text{d}z_{2}\phi(z_1) \phi(z_2) \xi(r(z_1, z_2, \theta), \hat{\textbf{r}}\cdot\hat{\textbf{l}}),
\end{equation}
which is a function of the angular separation defined through the relation,
\begin{equation}
    r(z_{1}, z_{2}, \theta) = \left(\chi(z_{1})^{2} + \chi(z_{2})^{2} - 2\chi(z_{1})\chi(z_{2})\cos{\theta}\right)^{1/2}.
\end{equation}
where $\phi(z) = N(z)D(z)$. The previously obtained power spectrum was computed at $z=0$, so $\phi(z)$ incorporates its evolution to a different redshift.

As mentioned before, the theoretical ACF with PNG shares similarities with the BAO template, but adding extra terms proportional to $\fnl$, to clarify this, we can consider that our PNG template is composed of a BAO-part and a $\fnl$-part, as follows,
\begin{equation}
    w(\theta)=w_{\rm BAO}(\theta)+w(\theta, \fnl),
\end{equation}
where $w_{\rm BAO}(\theta)$ is the BAO template used in \citetalias{descollaboration2021dark}, schematically given by,
\begin{equation}
    w_{\rm BAO}(\theta)\sim b^{2}w_{\scriptscriptstyle{b}}(\theta) + bf w_{\scriptscriptstyle{bf}}(\theta) + f^{2} w_{\scriptscriptstyle{f}}(\theta),
\end{equation}
where $w_{\scriptscriptstyle{b},\scriptscriptstyle{bf},\scriptscriptstyle{f}}(\theta)$ correspond to different ACF contributions arranged by their prefactors.
On the other hand, the $\fnl$-part involves the extra terms proportional to $\fnl$, in accordance with Eq.(\ref{eq:bias_fnl}), as follows,
\begin{equation}
    w(\theta, \fnl)\sim b\fnl w_{\rm A}(\theta) + \fnl^{2} w_{\rm B}(\theta)
\end{equation}
where $w_{\rm A,B}(\theta)$ involve the scale-dependent contributions of the ACF.
As a reminder of this discussion, we will extend the notation of our theoretical modelling to
\begin{equation}
    w(\theta)\rightarrow w_{\rm th}(\theta, \fnl),
\end{equation}
highlighting its dependence on $\fnl$. 

\begin{figure}
    \includegraphics[width=\columnwidth]{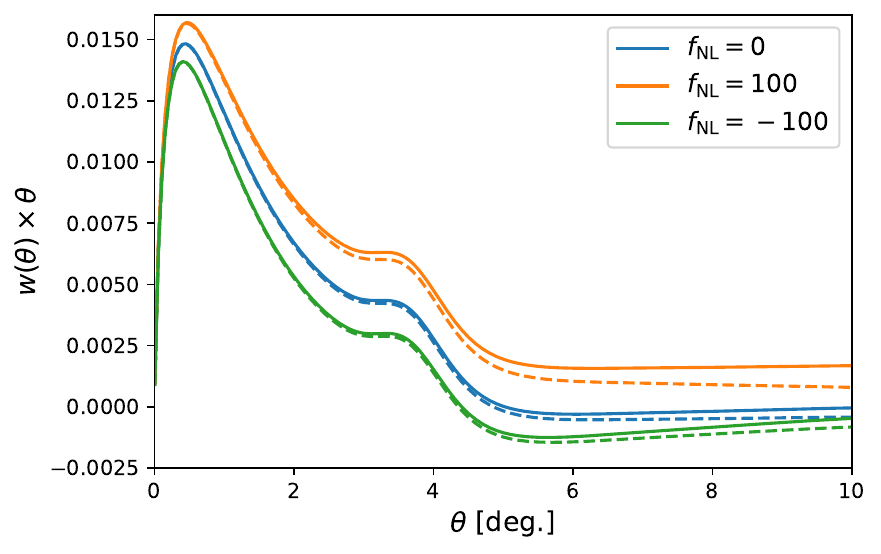}
    \caption{Theoretical angular correlation function with the scale-dependent bias for $\fnl=0$ (blue line), $\fnl=100$ (orange line), and $\fnl=-100$ (green line) for the first redshift bin using the ICE-COLA configuration as presented in Table \ref{tab:param}. The solid lines are without integral constraint. The dashed lines are with the integral constraint correction, as discussed in Section \ref{sec:ic}, computed using Eq.\ref{eq:ic_thetamax} with the ICE-COLA angular footprint.}
    \label{fig:fnl_acf}
\end{figure}

The behaviour of the angular correlation with PNG can be seen in Figure \ref{fig:fnl_acf}, where we compute the ACF using the BAO damped power spectrum, with linear bias and $N(z)$ from the first redshift bin of the ICE-COLA mocks.
As expected, we show that primordial non-Gaussianity induces a large-scale enhancement of clustering in the angular correlation function of galaxies due to the scale-dependent bias.
It can be noticed that the sharp feature in the power spectrum for $\fnl=-100$, produced due to the linear term in $\fnl$ (Eq.\ref{eq:bias_fnl}), has now translated into a small overall rising at scales around $\sim10$ degrees (solid green line in Figure \ref{fig:fnl_acf}).
This rising is due to the integration of the Fourier transform to compute the 2PCF.
As a preview of the upcoming section, we also show the integral constraint's effect on the theoretical model.
The main discussion of the upcoming section will be on how to compute the integral constraint correction and the effect on the ACF.

\section{Integral constraint and $\fnl$}\label{sec:ic}
In this section, we comment on how the excess of clustering at large scales, due to scale-dependent bias, on the theoretical angular correlation is suppressed by imposing that its integral over the survey volume needs to vanish. This condition is known as the integral constraint.

We discuss how the integral constraint arises from an observational point of view. We also remark on its dependence on $\fnl$ and show how to correct the theoretical template to incorporate its effect.

\subsection{Observational integral constraint}

\subsubsection{Integral constraint from the observed 2PCF}
Let us start with the statistical definition of the two-point correlation function for galaxies $\xi_{\rm obs}(r)$,
\begin{equation}
    \text{d}P=\bar{n}(1+\xi_{\rm obs}(r))\text{d}V,   \label{eq:cond_prob} 
\end{equation}
where $P$ is the probability of finding two objects within the volume $V$ separated by a distance $r$ \citep{1980lssu.book.....P} and $\bar{n}$ is the mean number density of galaxies in the Universe. If we integrate Eq.(\ref{eq:cond_prob}) over the volume of a survey, we find out that
\begin{equation}
    N_{g}=\bar{n}\int \text{d}V_{s}+\bar{n}\int \xi_{\rm obs}(r)\text{d}V_{s},
\end{equation}
where $N_{g}$ is the expected number of galaxies within the survey region and $V_{s}$ is the total volume of the survey. Since the expected number of galaxies within the survey volume is chosen to be obtained from the survey mean number density, we have the following,
\begin{equation}
    N_{g}=\bar{n} \int \text{d}V_{s}.   
\end{equation}
The previous equation implies a condition that needs to hold for the observed two-point correlation function of galaxies within the survey volume,
\begin{equation}\label{eq:ic_cond}
    \int \xi_{\rm obs}(r)\text{d}V_{s} = 0.    
\end{equation}
This is the integral constraint condition. We can re-write the integral constraint condition as follows,
\begin{equation}
    \int \xi_{\rm obs}(\r)\text{d}V_{s}=\int \text{d}^{3}\r \int \text{d}^{3}\r_{1} W(\r_{1})W(\r_{1}-\r)\xi(\r)=0,
    \label{eq:int_xi}
\end{equation}
where $W(\r)$ is the selection function for a volume-limited survey and $\r=\r_{1} - \r_{2}$.
    
The previous expression can be computed directly for a given survey selection function. The problem is that defining the volume of a survey is a difficult task. Instead, it is most common to construct a random catalogue of galaxies following the shape of the survey mask to model the survey volume as pair counts between the random catalogues.

As previously mentioned, the number count of galaxies within a homogeneous region can be computed as a volume integral of the selection function,
\begin{equation}
    N_{g}=\bar{n}\int \text{d}^{3}\r W(\r).
\end{equation}
Therefore, the number of random-random pair correlations, $RR(\r)$, can be computed as the correlation of the number of random objects within the limited region (see, e.g. \citet{2021Random}, and references therein),
\begin{equation}
    \begin{split}
        RR(\r)=\langle N_1 N_2\rangle=\bar{n}^{2}\int \text{d}^{3}\r_1 W(\r_1)W(\r_{1}-\r),   
    \end{split} \label{eq:rr}
\end{equation}
with $\r=\r_{1} - \r_{2}$. Using the previous equation, we can compute the volume integral over a window function, and inserting Eq.(\ref{eq:rr}) into Eq.(\ref{eq:int_xi}), we obtain the following,
\begin{equation}
    \begin{split}
        \int \xi_{\rm obs}(\r)\text{d}V_{s}=\frac{1}{\bar{n}^{2}}\sum_{\rm all\ pairs}RR(\r)\xi(\r), \label{eq:ic_xi_rr}
    \end{split}
\end{equation}
where now we sum over all the possible separations between galaxies within a limited survey size. This implies that the integral constraint condition, Eq.(\ref{eq:ic_cond}), can be written in terms of the $RR(\r)$ pairs, as follows,
\begin{equation}
    \begin{split}
        \sum_{\rm all\ pairs} RR(\r)\xi(\r)=0
    \end{split}
\end{equation}
where, for simplicity, the random-random pairs correlations can be obtained from random catalogues that follow the survey mask instead of using the analytic expression.
    
\subsubsection{Integral constraint in the observed ACF}
The previous procedure can be extended to the angular correlation function. The starting point is now the probability of finding two galaxies in a 2-dimensional projection of the sky separated by an angular separation $\theta$, as follows,
\begin{equation}
    \text{d}P =\bar{n}(1 + w_{\rm obs}(\theta))\text{d}\Omega,
\end{equation}
where $w_{\rm obs}(\theta)$ is the observed angular correlation function. 

This implies that the integral constraint can be extended to the angular correlation function in the same way as Eq.(\ref{eq:ic_xi_rr}),
\begin{equation}
    \int \text{d}\Omega_{1}  \int \text{d}\Omega_{2}W(\hat{\r}_1)W(\hat{\r}_2) w_{\rm obs}(\theta)=0,
\end{equation}
where $W(\hat{\r})$ is the angular selection function, and $\theta$ is the angle subtended by $\r_{1}$ and $\r_{2}$.

The calculation of the volume integral in the previous subsection can be extended to the sum of random-random angular pairs. This implies that we can compute the integral constraint for the angular correlation as follows,
\begin{equation}
    \sum_{\Omega} RR(\theta) w_{\rm obs}(\theta)=0, \label{eq:ic_w}
\end{equation}
where now the sum is over all the possible angular separations allowed by the survey mask. Also, as before, the random-random pairs correlation is obtained from the random catalogues. In practice, since we have a $w(\theta)$ for each redshift bin, this condition is applied to each of them individually.


\subsection{Theoretical integral constraint}\label{ssec:theo_ic}

Up to this point, we have only presented a condition that the correlation function needs to accomplish in limited surveys, and they certainly do for the usual observed correlation functions. A problem arises when we compare the theory with PNG to observational data.

\subsubsection{Gaussian case}

Let us start from a theoretical point of view without considering PNG. The matter power spectrum at large scales exhibits behaviour that goes as
\begin{equation}
    P_{\rm m}(k) \propto k^{n_s},
\end{equation}
where $n_{s}$ is close to 1. This implies that the matter power spectrum vanishes when $k\rightarrow0$, and since the power spectrum is related to the variance of the over-densities, this is an insight that the matter fluctuations of our Universe reach homogeneity at very large scales.

The vanishing of the matter power spectrum at large scales implies a condition to its configuration space counterpart, the 2PCF, which can be seen as follows,
\begin{equation} \label{eq:pk_0}
    P_{m}(k=0) = \int \xi(\r)  \text{d}^{3}\r = 4 \pi \int_{0}^{\infty} \xi(r) r^2 \text{d}r = 0. 
\end{equation}
This is the integral constraint condition presented in Eq.(\ref{eq:int_xi}) but now coming from a purely theoretical perspective.

Without the effect of $\fnl$, this same condition is expected to hold for the linear galaxy power spectrum since a linear bias relates both power spectra, and there is no change in the shape of the power spectrum. Hence, in the case of an ideal homogeneous infinite survey, the theoretical model already satisfies the observational integral constraint. When the effect of the window function becomes more pronounced (due to either strong inhomogeneities in the randoms or small explored volumes), we will need to adjust the theory to fulfil the IC condition (see subsection \ref{ssec:ic_correction}).

\subsubsection{Integral constraint in the presence of PNG}

The situation now changes in the presence of PNG. The scale-dependent bias between the galaxies and matter overdensities will modify the shape of the galaxy power spectrum introducing a $1/k^{2}$ correction to the matter power spectrum that depends on $\fnl$, as described in Eq.( \ref{eq:alpha}).
The scale-dependent bias will generate an enhancement of the galaxy power spectrum at large scales ($k\ll k_{\rm eq}$) with the following divergent behaviour,
\begin{eqnarray}
    P_{g}(k\to0,\fnl) \sim \Bigg( \fnl \frac{b-1}{k^2} \Bigg)^2 \cdot k^{n_s} \to \infty 
\end{eqnarray}
where $P_{g}(k)$ is the galaxy power spectrum. This divergence that the volume integral over the 2PCF (Eq.(\ref{eq:pk_0})) will diverge for this case. As a side note, since for our modelling, we integrate numerically, the previously mentioned divergence will turn into a large (but finite) number that could depend on the integration method or resolution.
Since the ACF is an integral of the 3D 2PCF (Eq.(\ref{eq:acf_fnl})), $w(\theta)$ will have a divergence proportional to $\fnl^{2}$.

The discussion of this section tells us that, even if we have an infinite homogeneous survey with a negligible window function effect, the integral constraint condition will not be fulfilled for the case of $\fnl\neq 0$. Additionally, the theoretical model will contain an arbitrary additive constant that depends on $\fnl^2$. This dependence will bias any results when using this model to constrain $\fnl$.
This remarks the importance of the integral constraint condition when dealing with PNG, implying that we need correct our modelling to consider this issue. 

As a verification of the issue, in appendix \ref{ap:acf_theory}, we show an analytical example that illustrates how the integral constraint condition looks for a simplified theoretical 2-point correlation function in the presence of PNG. We show explicitly that the integral of the 2PCF diverges at large scales and is proportional to $\fnl^{2}$, implying that imposing the observational integral constraint condition is very important when dealing with PNG analysis. 

\subsection{Integral constraint correction}\label{ssec:ic_correction}
To surpass the problem described in the previous subsection, we define an integral constraint-corrected theoretical angular correlation function,
\begin{equation}\label{eq:acf_ic}
    w^{\rm IC}(\theta, \fnl)=w_{\rm th}(\theta, \fnl) - I(\fnl),
\end{equation}
where $I(\fnl)$ parametrize deviations from the observed integral constraint condition (Eq.\ref{eq:ic_w}) as follows,
\begin{equation}
    \sum_{\Omega} RR(\theta) w^{\rm IC}(\theta, \fnl)=0 \label{eq:ic_w_corr}.
\end{equation}
This implies that the integral constraint correction, $I(\fnl)$, is given by:
\begin{equation}
    I(\fnl)=\frac{\sum^{\theta_{\rm lim}} RR(\theta) w_{\rm th}(\theta, \fnl)}{\sum^{\theta_{\rm lim}} RR(\theta)}. \label{eq:ic_thetamax}  
\end{equation}
where $\theta_{\rm lim}$ is the maximum limit angular separation allowed for the survey angular window.
The effect of the integral constraint in the context of PNG has been previously addressed in \citet{Ross2013,mueller2021clustering} for the power spectrum and in \citet{Ross2013} for the 2PCF. The novelty of this work is to present a detailed analysis of its effect on the ACF and show its importance when dealing with PNG simulations, as we will show in Section \ref{sec:test_goliat}.

\section{Simulations}\label{sec:sims}
In this section, we present the simulations that we used for testing the theoretical modelling and the validation of the $\fnl$ measurements.

\subsection{GOLIAT-PNG}\label{ssec:goliat}

In order to test our analysis pipeline, we first consider the use of simulations with Primordial non-Gaussianity included. Whereas many tests can be done with Gaussian initial conditions (see Section \ref{sec:opt_cola}), there are validation steps that require PNG mocks to show the validity of the pipeline. In particular, in this work, only when fitting PNG mocks can we realise the paramount importance of including the integral constraint. 

The GOLIAT-PNG suite \citep{2023MNRAS.519.3706A} consists of a series of $N$-body simulations with $\Lambda$CDM + local PNG cosmology with $\Omega_m=0.27$, $\Omega_{b}=0.044$, $h=0.7$, $n_s=0.96$, $\sigma_8=0.8$, and three values for PNG: $f_{\rm NL}=-100,0,+100$. A summary of the cosmological parameters and fiducial values used is presented in the first part of Table \ref{tab:param}. The simulations have a box size of $L=1{\rm Gpc}/h$. The initial conditions are set at $z=32$ with second-order Lagrangian perturbation theory using the public code \textsc{2LPTic} \footnote{\url{cosmo.nyu.edu/roman/2LPT} \citep{2LPTic,2LPT-PNG}} and evolved to $z=1$ with \textsc{Gadget2} \footnote{\url{https://wwwmpa.mpa-garching.mpg.de/gadget/} \citep{Gadget2}}. Subsequently, the $z=1$ dark matter snapshots are run through the Amiga Halo Finder \footnote{\url{http://popia.ft.uam.es/AHF/} \citep{AHF}} to construct the halo catalogues with a minimum of 10 particles, which yield $M_h\sim 5\cdot 10^{12}M_{\odot}$ as the halo mass resolution.

Also, for the \goliat simulations, it was found that $p=0.90$ for $\fnl=100$ \citep{2023MNRAS.519.3706A}, and $p=0.92$ for $\fnl=-100$, when measuring their real space power spectra, and we will consider this when measuring $\fnl$ from these mocks. 

Another particularity of these simulations is that the initial conditions are run with the \textit{Fixed \& Paired} initial conditions \citep{Angulo16} aimed at reducing the sample variance of the ensemble average of the 2-point functions measured from these simulations. In the context of PNG, this technique is validated in \citet{2023MNRAS.519.3706A}, and we refer the reader there for further details of the GOLIAT-PNG suite. 
We use 41 pairs of simulations for each value of $\fnl$. 

Finally, we transform those mocks from the cubic box into observable coordinates $\{$ra,dec,$z\}$ by setting an observer $1556 {\rm Mpc}/h$ away from the centre of one of the faces of the box. This transformation allows us to have a mock survey with a circular semi-aperture of 11.2 deg, covering an area of roughly 396 $\text{deg}^{2}$, and a redshift range of $0.6<z<1.1$, the shape and size of the mask can be seen in Figure \ref{fig:mask}. We further split the mocks into five redshift bins between 0.6 and 1.1 with $\Delta z = 0.1$. This, together with a constant number density of halos, give the redshift distribution $N(z)$ shown in Figure \ref{fig:nz}.
However, we note that we do not introduce any redshift space distortions, redshift error, HOD model, or even temporal evolution. We built everything from the halo catalogue at the comoving output at a redshift of $z=1$ and a fixed halo mass threshold. This implies that we fix $D(z=1)$ in Eq.\ref{eq:acf_fnl} when using the \goliat mocks. We also consider three different rotations (one per Cartesian axis) for constructing the mocks, eventually resulting in a total of 246 mocks for each value of $\fnl$.

\textcolor{purple}{}

\subsection{ICE-COLA}\label{ssec:cola}

The ICE-COLA mocks \citep{DES:2021fie} are the second set of simulations we count on for analysing and validating our methods. This set of 1952 mock galaxy catalogues is designed to mimic the DES Year 3 BAO sample \citep{rosell2021dark} over its full photometric redshift range $0.6<z<1.1$, which we split again into five redshift bins. We refer the interested reader to \cite{DES:2021fie} for further details and highlight here only the basic features of the ICE-COLA mocks.
 
 A total number of 488 fast $N$-body simulations of full-sky light cones generated by following the ICE-COLA code \citet{ice-cola} are used. This code is based on the COmoving Lagrangian Acceleration (COLA) method, which solves for the evolution of the matter density field using second-order Lagrangian Perturbation Theory (2LPT) combined with a Particle-Mesh (PM). The simulations use $2048^3$ particles in a box of the size of 1536 Mpc h$^{-1}$ and assume a cosmology consistent with the best-fit of WMAP five-year data \citep{Dunkley09}. This means compatible with a flat $\Lambda$CDM model with $\Omega_m=0.25$, $\Omega_{\Lambda}=0.75$, $\Omega_b=0.044$, $n_s=0.95$, $\sigma_8=0.8$, $h=0.7$, and $\fnl=0$. A summary of the cosmological parameters and fiducial values used is presented in the second part of Table \ref{tab:param}.
 
\begin{figure}
    \includegraphics[width=\columnwidth]{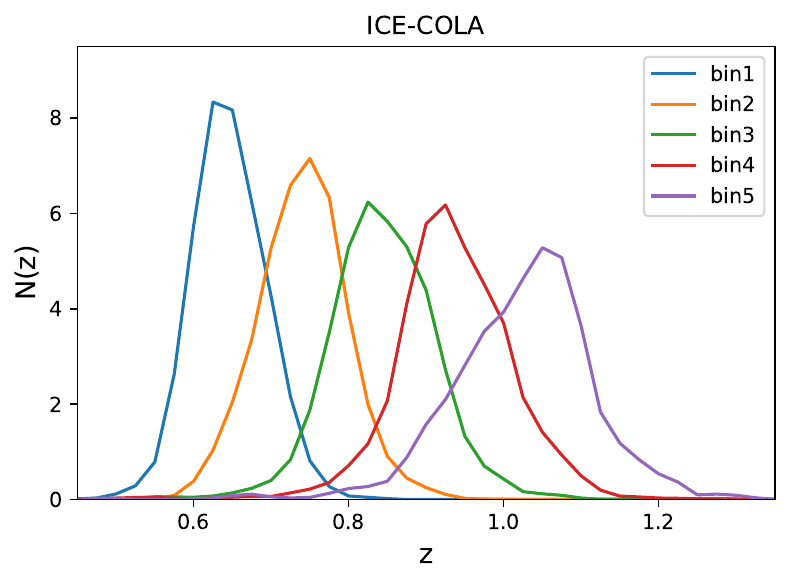}
    \includegraphics[width=\columnwidth]{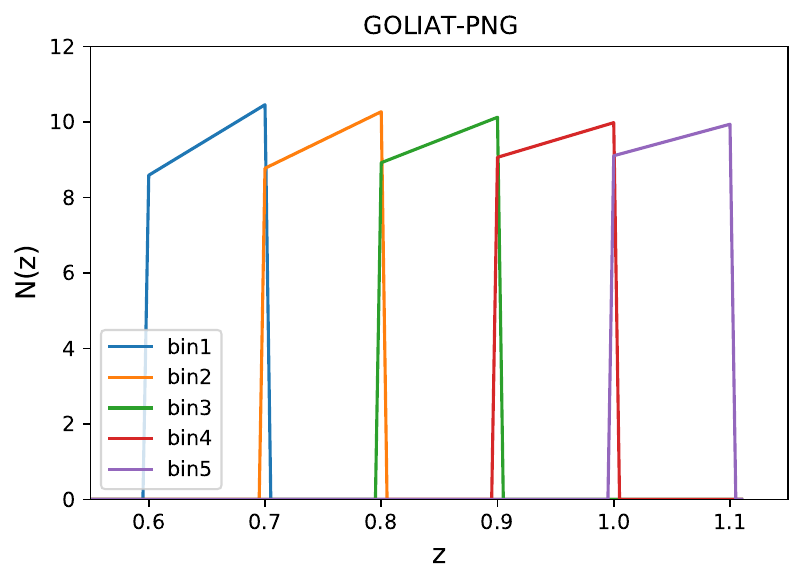}
    \caption{$N(z)$ distribution as a function of redshift for each redshift bin for the ICE-COLA mocks (top) and for the GOLIAT-PNG mocks (bottom). 
    We remark that the \goliat simulations do not have photo-z errors included, implying that they do not represent a realistic $N(z)$ distribution, in contrast to the ICE-COLA mocks.}
    \label{fig:nz}
\end{figure}

A hybrid halo occupation distribution - halo abundance matching model is used to populate halos with galaxies. Also, automatic calibration is run to match the basic characteristics of the DES Y3 BAO sample: the observed abundance of galaxies as a function of photometric redshift (Figure \ref{fig:nz}), the distribution of photometric redshift errors, and the clustering amplitude on scales smaller than those used for BAO measurements.

Finally, four footprint masks corresponding to the DES Y3 BAO sample are placed on each full-sky light cone simulation to reach the final set of 1952 ICE-COLA mocks. In Figure \ref{fig:mask}, we can see the shape of the mask followed by one footprint.

\begin{figure}\label{fig:mask}
    \includegraphics[width=\columnwidth]{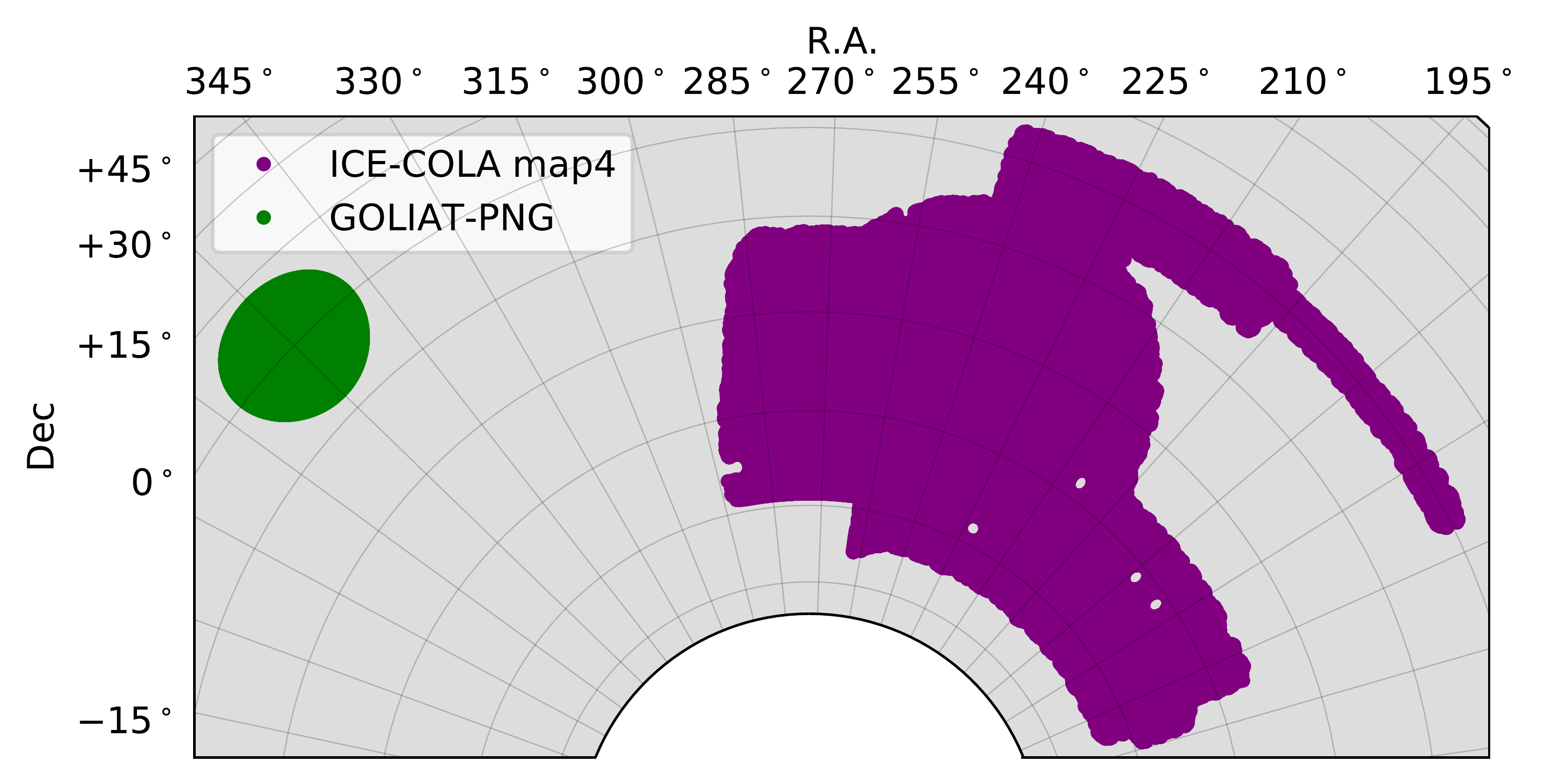}
    \caption{Comparison of the footprint of the used simulations. In purple, we show the mask for one map of the ICE-COLA simulations. In green, we show the mask for the \goliat simulation.}
    \label{fig:mask}
\end{figure}

\section{Analysis tools}\label{sec:tools}
This section presents the statistical tools used to measure the $\fnl$ parameter using the theoretical template presented in Section \ref{sec:theory}. 

\subsection{Angular correlation function measurements}\label{sec:acfM}
The angular correlations are measured using CUTE \citep{2012arXiv1210.1833A}, which computes the ACF following the Landay-Szalay estimator (\citet{1993ApJ...412...64L}),
\begin{equation}
    w_{\rm obs}(\theta) = \frac{DD(\theta)-2DR(\theta)+ RR(\theta)}{RR(\theta)},
\end{equation}
where $DD(\theta)$, $DR(\theta)$ and $RR(\theta)$ are the number counts of pairs of galaxies for the data-data, data-random, and random-random catalogues, respectively. To obtain the random-random pairs, we create random catalogues with 20 times more objects than the simulation sample that follow the angular mask of the simulations for \goliat and ICE-COLA mocks. The random-random pairs are obtained as an output from CUTE.

As mentioned in subsection \ref{ssec:ic_correction}, one of the key elements in the integral constraint correction is the random-random pairs that account for the survey volume. Because of this, we need to compute at least one $RR(\theta)$ correlation for both \goliat and ICE-COLA going up to the maximum angular separation allowed for each survey mask. That is 22 degrees for the \goliat simulations and 88 degrees for ICE-COLA simulations. 

\subsection{Covariance} \label{ssec:cov}

Our default setup for the covariance matrix uses the Cosmolike code \citep{cosmolike,cosmolike2020,cosmolike_curvedsky}  to estimate the covariance analytically.   
Following \cite{Crocce2011}, the real space covariance of the angular correlation function $w(\theta)$ at angles $\theta_i$ and $\theta_j$ is related to the covariance of the angular power spectrum $C(C_{\ell}, C_{\ell'})$ by
\begin{equation}
C(\theta_{i}, \theta_{j}) = \sum_{\ell, \, \ell'}\dfrac{(2\ell +1)(2\ell'+1)}{(4\pi)^2}\overline{P_{\ell}}(\theta_i)\overline{P_{\ell'}}(\theta_j)  C(C_{\ell}, C_{\ell'}),
\end{equation}
where $\overline{P_{\ell}}(\theta)$ are the Legendre polynomials averaged over each angular bin and $C(C_{\ell}, C_{\ell'})$, under the Gaussian approximation, is given by \citep{Crocce2011,cosmolike}
\begin{equation}
C(C_{\ell}, C_{\ell'}) =  \dfrac{2\delta_{\ell \ell'}}{f_{\rm sky}(2\ell + 1)}\left(C_{\ell'} + \frac{1}{n_g}\right)^2,
\label{eq:cov-c_ell}
\end{equation}
where $\delta$ is the Kronecker delta function, $n_g$ is the number density of galaxies per steradian, and $f_{\mathrm{sky}}$ is the observed sky fraction used to account for partial-sky surveys. We include redshift space distortions through the $C_{\ell}$'s of the expression above \eqref{eq:cov-c_ell}, except when analysing the \goliat mocks, as they do not include that.
In addition, following \cite{cosmolike_mask}, we correct the shot-noise contribution to the covariance (the term $\propto1/n_g$) by considering the effect of the survey geometry on the number of galaxies in each angular bin. 
We ignore non-Gaussian terms in the covariance estimation for simplicity, following \citetalias{descollaboration2021dark}, where it was tested that including those terms did not impact the results.
See \citetalias{descollaboration2021dark} and \citet{DES:2021fie} for the validation of this analytical covariance matrix (with $\theta_{\mathrm{max}}=5$ deg) against two sets of simulations: ICE-COLA and FLASK lognormal mocks \citep{Xavier:2016elr}.

Notice that we do not include $\fnl$ in our covariance since it is customary in this kind of analysis to fix the cosmology and then look for deviations. In the case of detection, we should modify the covariance and include the $\fnl$ parameter.

We also consider using the ICE-COLA covariance constructed from the mocks, given by:
\begin{eqnarray}
    C(\theta_{i}, \theta_{j}) = \frac{1}{\rm N_{\rm m} -1}\sum_{n=1}^{\rm N_{\rm m}}\left(w^{n}(\theta_{i})-\bar{w}(\theta_{i}) \right)\left(w^{n}(\theta_{j})-\bar{w}(\theta_{j}) \right)
\end{eqnarray}
where $\rm N_{\rm m}$ is the number of mocks, $w^{n}(\theta)$ is the ACF for the n-mock, and $\bar{w}(\theta)$ is the mean ACF from the mocks.
However, it was shown in \citet{DES:2021fie} that, due to a large number of simulated boxes used to equal the volume of the DES Y3 BAO sample, a replication of halos were produced, introducing a spurious correlation among the measured ACF. This induced a high degree of correlation of non-adjacent redshift bins in the covariance.
For this reason, the default setup of using Cosmolike covariance was preferred \citepalias{descollaboration2021dark}. As a double check, in subsection \ref{ssec:cov_comp}, we compare the impact of changing the covariance when measuring $\fnl$.

\subsection{Parameter inference}\label{ssec:mcmc}
In order to measure the parameters, we perform a Bayesian parameter inference based on the log-likelihood analysis assuming a Gaussian likelihood, as follows,
\begin{equation}
    \log(\mathcal{L}(\p))\propto -\frac{\chi^{2}(\p)}{2}
\end{equation}
where the $\chi^{2}$ is given by,
\begin{equation}
    \chi^{2}(\p) = (M(\p) - D)^{T} C^{-1} (M(\p) - D)
\end{equation}
where $\p$ represents the free parameters from our theory we want to estimate, $C^{-1}$ is the inverse of the covariance matrix presented in subsection \ref{ssec:cov}, and $M$ and $D$ are the theoretical model and the data vector, respectively.

Since the galaxy sample for the simulations is divided into five redshift bins, we perform a joint sampling of the likelihood to consider covariance between bins. The joint data vector $\textbf{D}$ is given by,
\begin{equation}
    \textbf{D} = [w_{\rm obs}^{1}(\theta), w_{\rm obs}^{2}(\theta), w_{\rm obs}^{3}(\theta), w_{\rm obs}^{4}(\theta), w_{\rm obs}^{5}(\theta)],
\end{equation}
where the superscript represents the redshift bin from which the ACF is obtained. We repeat the same procedure for the theoretical model, where $\textbf M(\p)$ is the theory vector as a function of the free parameters for each redshift bin, as follows,
\begin{equation}
    \textbf{M}(\p) = [w_{\rm th}^{1}(\theta, \p), w_{\rm th}^{2}(\theta, \p), w_{\rm th}^{3}(\theta, \p), w_{\rm th}^{4}(\theta, \p), w_{\rm th}^{5}(\theta, \p)].
\end{equation}

We perform an MCMC sampling of the likelihood function using COBAYA \citep{2021Cobaya} to estimate the posterior distributions of the free parameters in our pipeline.

Table \ref{tab:param} presents the free parameters considered for our analysis and their respective fiducial values and priors.
Depending on the analysis, the integral constraint could be considered as a free parameter (IC-MARG) or fixed to its theoretical value (IC-FIXED) given by Eq.(\ref{eq:ic_thetamax}). This will be stated for each test considered. Notice that we are not including other free cosmological parameters in the likelihood, which is customary for this kind of analysis, since adding other cosmological parameters will lose the constraints on $\fnl$.

\begin{table}
    \centering
    \caption{Summary of the fixed cosmological parameters and the free measured parameters with the priors considered. The squared brackets represent flat priors.}
    \label{tab:param}
    \begin{tabular}{lcccc} 
	    \hline \hline & \textbf{GOLIAT-PNG} & & \\ \hline \hline
        Parameter & Fiducial  & Prior \\ \hline
        $\Omega_{m}$ & $0.27$ & -- \\
        $\Omega_{\Lambda}$ & $0.73$ & -- \\
        $\Omega_{b}$ & $0.044$ & -- \\
        $n_{s}$ & $0.96$ & -- \\
        $\sigma_{8}$ & $0.8$ & -- \\
        $h$ & $0.7$ & -- \\
	    $\fnl$ & $-100, 100$ & $[-700, 700]$\\
	    Linear bias $b$ & $2.35$ & $[1, 3]$ \\
	    Integral constraint $I_{i}$ & - & $[-0.1, 0.1]$\\
        Footprint area ($\text{deg}^2$) & 396.06 & \\
	    $z_{\rm mean}$ & $1$ & -- \\
     \hline \hline & \textbf{ICE-COLA} & & \\ \hline \hline
        Parameter & Fiducial  & Prior \\ \hline
        $\Omega_{m}$ & $0.25$ & -- \\
        $\Omega_{\Lambda}$ & $0.75$ & -- \\
        $\Omega_{b}$ & $0.044$ & -- \\
        $n_{s}$ & $0.95$ & -- \\
        $\sigma_{8}$ & $0.8$ & -- \\
        $h$ & $0.7$ & -- \\
	    $\fnl$ & 0 & $[-500, 500]$\\
	    Linear bias $b_{i}$ & $1.60$, $1.60$, $1.68$, $1.82$, $2.02$ & $[1, 3]$ \\
	    Integral constraint $I_{i}$ & - & $[-0.1, 0.1]$\\
        Footprint area ($\text{deg}^2$) & 4108.47 & \\
	    $z_{\rm mean}$ & $0.65$, $0.74$, $0.84$, $0.94$, $1.02$ & -- \\ \hline
    \end{tabular}
\end{table}


\section{Tests with non-Gaussian mocks}\label{sec:test_goliat}
This section tests the pipeline over the \goliat simulations with non-Gaussian initial conditions. For these simulations, the theoretical template is obtained from a linear power spectrum without considering BAO damping and without RSD modelling since the simulations do not include RSD.
The goal of this section is twofold: First, we want to recover the fiducial value of $\fnl$ for the non-Gaussian simulations. Second, we want to highlight the importance of the integral constraint.

\subsection{Effect of the Integral constraint on \goliat mocks}\label{ssec:IC_thetamax}
In Section \ref{sec:ic}, we presented the integral constraint as one of the key elements that need to be included in the theory. In this section, we show its effect in the simulations with non-Gaussian initial conditions.
\begin{figure}
    \includegraphics[width=\columnwidth]{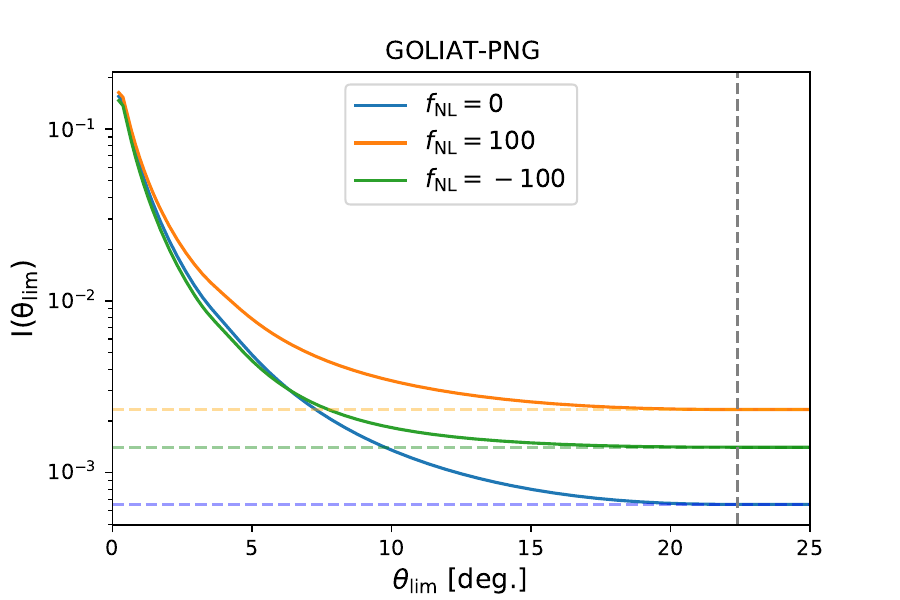}
    \caption{Integral constraint as a function of the upper limit angular separation, $\theta_{\rm lim}$, for the \goliat simulations. The blue line is the integral constraint using the theoretical ACF with $\fnl=0$. The same is repeated for the orange and green lines but for the cases of $\fnl=100$ and $\fnl=-100$, respectively. The grey dotted line is the limit angular aperture of the angular mask of the simulations.}
    \label{fig:IC_nbody1}
\end{figure}

In Figure \ref{fig:IC_nbody1},  we compute the integral constraint, as presented in Eq.(\ref{eq:ic_thetamax}), for the \goliat simulations but changing the limit angular separation, $\theta_{\rm lim}$, truncating the sum.
We use this to test the need to consider the full volume of the survey when computing the integral constraint.
As described in subsection \ref{ssec:goliat}, the maximum circular semi-aperture of the $\goliat$ simulation mask is about 11.2 degrees, implying that the maximum allowed angular separation is about $\theta_{\rm lim}\sim22$ degrees (vertical grey dotted line in Figure \ref{fig:IC_nbody1}).

As expected, given the discussion in Section \ref{sec:ic}, the integral constraint reaches its value when it is summed up to the maximum angular separation allowed for the simulation mask to consider the whole survey volume.
In other words, even though we can compare the theory and the data up to some maximum angular separation $\theta_{\rm max}$, we still need the random-random correlation up to the limit scale of the simulation ($\theta_{\rm lim}\sim 22$ deg). We see that the integral constraint's value does not converge earlier than that.
We repeat this conclusion for the ICE-COLA simulations, where the measurements are made up to $\theta_{\rm max}=20$ degrees, but the integral constraint is obtained from random-random pairs measured up to $\theta_{\rm lim}\sim 88$ degrees.

From the previous figure, we can also notice the explicit dependence of the integral constraint on $\fnl$. For $\fnl=0$, it has a smaller value in comparison with $\fnl=100$ or $\fnl=-100$. This supports the previous discussion from subsection \ref{ssec:theo_ic} about the importance of the integral constraint when looking for $\fnl$.

\begin{figure}
    \includegraphics[width=\columnwidth]{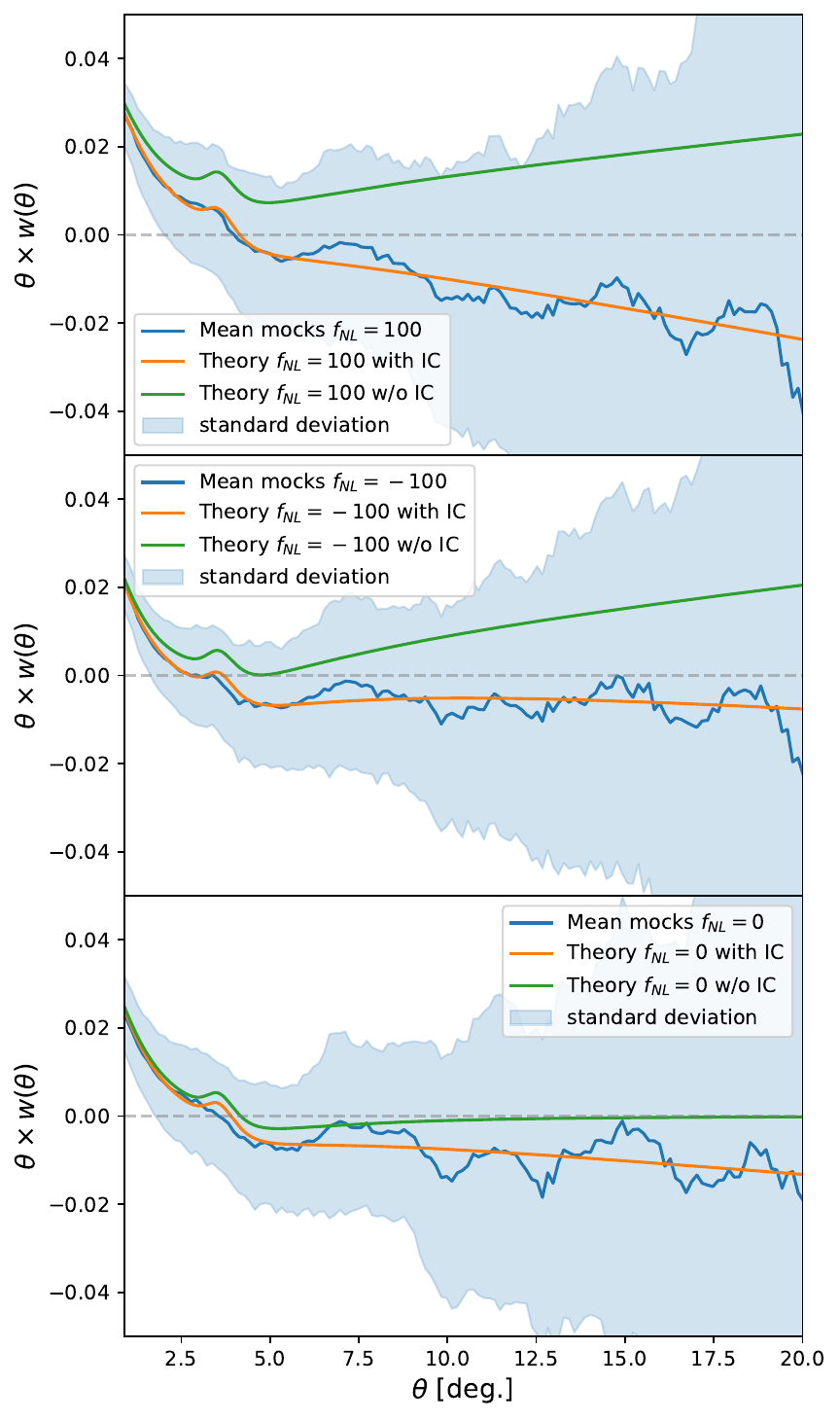}
    \caption{Comparison of the theoretical ACF against the mean \goliat mocks for the first redshift bin ($0.6<z<0.7$) and different $\fnl$ values. The blue line is the mean of the mocks, and the shaded area is given by its standard deviation. The orange line is the theoretical ACF with the integral constraint. The green line is the theoretical ACF without the integral constraint.}
    \label{fig:acf_goliat100}
\end{figure}

The previously computed integral constraint can be included in the theory as in Eq.(\ref{eq:acf_ic}). This is shown in Figure \ref{fig:acf_goliat100}, where we compare the theoretical ACF with and without the integral constraint against the mean of the \goliat mocks. The ACF is shown for the first redshift bin with the errors obtained from the standard deviations of the mocks.
Figure \ref{fig:acf_goliat100} serves as a visual guide of the integral constraint's effect in the theoretical modelling. The integral constraint correction appears to have an effect that could help avoid biased values for $\fnl$.
The actual impact of this on the measurement of $\fnl$ is the main topic of the following subsection. 

\subsection{Results for \goliat mocks}\label{ssec:goliat_results}

We use the parameter inference method, described in subsection \ref{ssec:mcmc}, to put constraints on both the linear bias and $\fnl$.
We construct the data vector for each mock by combining the ACF of each redshift bin for the $\fnl=-100$ and $\fnl=100$ simulations.
We use the scale configuration given in the first section of Table \ref{tab:fiducial_conf}. The scale choice will be justified in the next section when we test the robustness of the pipeline.
 
Since each mock is independent of the other, we can compute a joint posterior distribution by multiplying the posteriors of $\fnl$ and $b$ of each \goliat mock. The advantage of this method is that the joint posterior gives us a good estimate of how biased the best-fit values of $\fnl$ are with respect to the fiducial.
We compare fixing the IC, as computed using Eq.(\ref{eq:ic_thetamax}), against not using it and against leaving it as a nuisance parameter. The priors for the parameters used in the measurement are in Table \ref{tab:param}. For the case of $\fnl=100$ simulations, four mocks were discarded due to incompatibilities in the measurements of $\fnl$, giving highly biased values and complicating the computation of the joint posterior.

\begin{figure*}
    \includegraphics[width=\columnwidth]{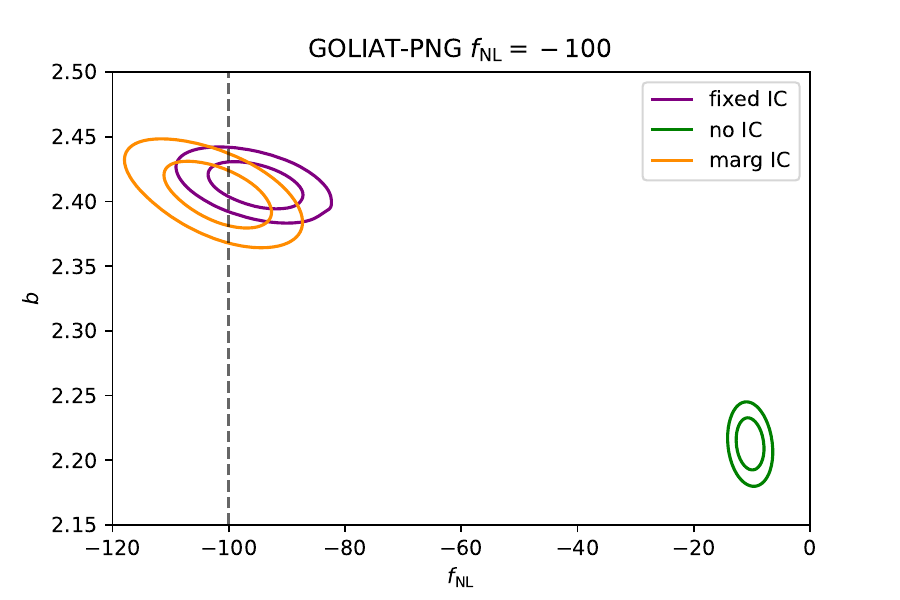}
    \includegraphics[width=\columnwidth]{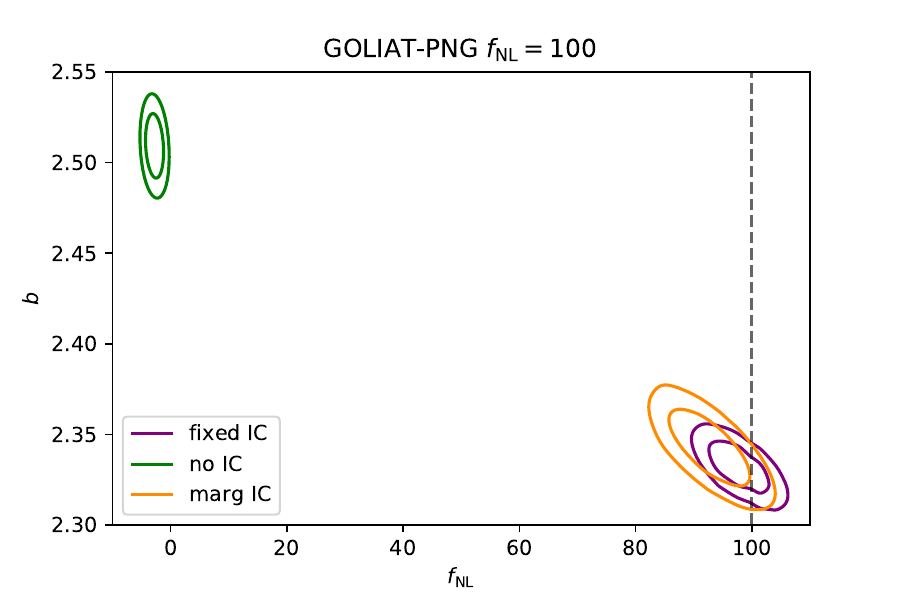}
    \caption{Marginalized one and two-sigma contours for $\fnl$ and the linear bias $b$ obtained from the joint posterior of the 246 \goliat simulations (of $\sim 400\rm{deg}^2$ each). Note that the error is expected to be $\sim 16$ times larger for a single realisation. The left panel is for the $\fnl=-100$ simulation and the right panel is for the $\fnl=100$ simulation. The purple contours are with the integral constraint fixed to its theoretical value given by Eq.(\ref{eq:ic_thetamax}). The green contours are without considering any integral constraint correction. The orange contours consider the IC as a nuisance parameter and marginalising it. The vertical dashed line represents the fiducial value of $\fnl$ for each set of simulations.}
    \label{fig:triangle_fnl100}
\end{figure*}

\begin{table}
    \centering
    \caption{Summary of the results of measuring $\fnl$ from both \goliat simulations (of $\sim 400\rm{deg}^2$ each). The best-fit values are obtained from the maximum of the joint posterior of the 246 mocks, and the errors are at $1\sigma$. Note that the error is expected to be $\sim 16$ times larger for a single realisation.}
    \label{tab:comp_test_goliat}
    \begin{tabular}{lc} 
	    \hline \hline \textbf{\goliat} & \\
	    \hline \hline
	     & Joint posterior \\ \hline \hline
	    $\fnl=100$ \\ \hline
	    NO-IC & $-2.8\pm1.0$ \\
	    \textbf{IC-FIXED} & $97.4\pm3.5$ \\
	    IC-MARG & $92.2\pm4.6$ \\
	    \hline
	    $\fnl=-100$ \\ \hline
	    NO-IC & $-10.3\pm1.5$ \\
	    \textbf{IC-FIXED} & $-95.2\pm 5.4$ \\
	    IC-MARG & $-101.5\pm6.5$ \\
	    \hline
    \end{tabular}
\end{table}

We present one of the main results of this work in Figure \ref{fig:triangle_fnl100}, showing the contours obtained from the joint posterior of all \goliat simulation with $\fnl=100$ and $\fnl=-100$.
We show that by fixing the integral constraint to the value given by Eq.(\ref{eq:ic_thetamax}), we can recover the fiducial values of $\fnl$ within 1$\sigma$. We also notice that for the case of not using the integral constraint, we obtain very biased values for $\fnl$, closer to $\fnl=0$. The figure also shows that when considering the integral constraint as a nuisance parameter and marginalising it, we also recover the correct values for $\fnl$.
With the previous results, we prove the importance of the integral constraint.


The summary of contours is presented in Table \ref{tab:comp_test_goliat}, where we show the measured values of $\fnl$ for the two kinds of \goliat simulations. The best-fit values of $\fnl$ are obtained from the maximum of the joint posterior distribution of all mocks, with the errors obtained from the $68\%$ confidence region.
We clarify that the uncertainty presented in Table \ref{tab:comp_test_goliat} corresponds to the combination of all mocks.
This implies that the uncertainty would be $\sim 16$ times larger for a survey with the properties of the \goliat mocks, making the uncertainty and the offset very similar $\Delta \fnl \sim \sigma \sim 100$. We also note that the relatively small footprint of \goliat ($\sim 400 \rm{deg}^2$) makes the effect of the IC stronger. We will reexamine this for a DES-like scenario in Section \ref{ssec:ic_cola}.

\begin{table}
    \centering
    \caption{Comparison between theoretical integral constraint versus marginalised values for \goliat simulations. IC theory is computed using the theoretical value given by Eq.(\ref{eq:ic_thetamax}). IC marginalised are obtained as the mean of the marginalised posterior. The errors are at $1\sigma$ on the ensemble average of 246 mocks.}
    \label{tab:ic_comp_goliat}
    \begin{tabular}{lcc} 
	    \hline \hline 
     \multicolumn{3}{|c|}{\textbf{\goliat}} \\
	    \hline \hline
	 Redshift bin & IC theory & IC marginalized  \\ \hline \hline
  $0.6 < z < 0.7$ & 0.00220 & $0.00247 \pm 0.00013$ \\ 
  $0.7 < z < 0.8$ & 0.00202 & $0.00208 \pm 0.00012$ \\ 
  $0.8 < z < 0.9$ & 0.00188 & $0.00212 \pm 0.00012$ \\ 
  $0.9 < z < 1.0$ & 0.00178 & $0.00203 \pm 0.00011$ \\ 
  $1.0 < z < 1.1$ & 0.00171 & $0.00184 \pm 0.00011$ \\ \hline
    \end{tabular}
\end{table}

A natural question appears when we see the results for the case of IC-MARG. Can the marginalised IC case recover the theoretical values given by Eq.(\ref{eq:ic_thetamax})? In Table \ref{tab:ic_comp_goliat}, we compare the IC values for both theoretical and marginalised, along with the $1\sigma$ errors for the marginalised case measured over the mean of the mocks. From these results, we can notice two things. First, we found reasonable compatible values for the integral constraint within $\sim 2\sigma$. Secondly, we show that the methods can detect the integral constraint at high significance.

\begin{figure*}
    \includegraphics[width=\textwidth]{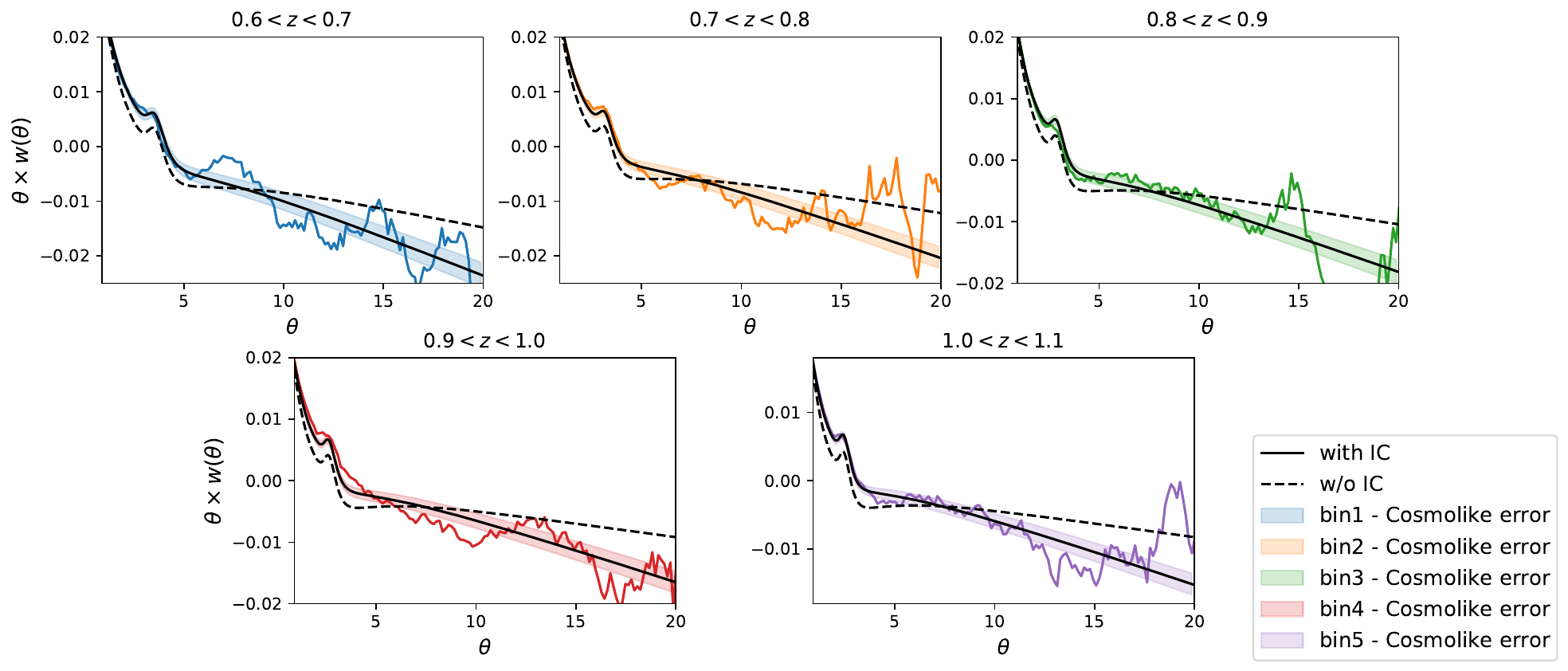}
    \caption{Comparison between the theoretical angular correlation versus the mean ACF of the \goliat mocks with $\fnl=100$, for each redshift bin. The solid-coloured lines are the mean of the ACF from the mocks. The shaded areas are obtained from the diagonal of the reduced theoretical covariance. The solid black lines are theoretical ACF with integral constraint, where $\fnl$ and $b_g$  are obtained from the mean of the joint posterior distribution presented in purple in Figure \ref{fig:triangle_fnl100}. The dashed black lines are the theoretical ACF without integral constraint and $\fnl$ and $b_g$ obtained from the mean of the joint posterior distribution presented in the green lines of Figure \ref{fig:triangle_fnl100}.}
    \label{fig:acf_fnl100}
\end{figure*}

In Figure \ref{fig:acf_fnl100}, we compare the mean of the \goliat $\fnl=100$ mocks versus the theoretical ACF (for IC-FIXED) using the best-fit results with and without the integral constraint for each redshift bin. The figure shows how the integral constraint improves the agreement of the theoretical template and the observed ACF for each redshift bin. Nevertheless, we found no considerable difference in $\chi^{2}$ of the measurement over the individual mocks when considering or not the integral constraint in the theoretical template.
The showed errors, in this case, are obtained from the Cosmolike covariance, described in subsection \ref{ssec:cov}, but divided by the number of mocks, in contrast with the errors presented in Figure \ref{fig:acf_goliat100}. 

For the case of NO-IC, we notice that for both simulations, we obtain biased small negative values of $\fnl$. As mentioned by the end of Section \ref{ssec:acf_png}, for large negative values of $\fnl$ (without considering IC), there is a positive correlation function at large scales (see, for example, middle panel of Fig. \ref{fig:acf_goliat100}). Since the measured angular correlation function shows a negative correlation at large scales (due to the observational integral constraint), the model prefers small negative $\fnl$ values to compensate for the lack of IC in the theoretical model (see, e.g. Fig. \ref{fig:acf_fnl100})


As mentioned in subsection \ref{ssec:theo_ic}, the effect of the integral constraint is stronger for non-Gaussian simulations due to its dependence on $\fnl$. Nevertheless, in the next section, we will show that it can also help avoid slightly biased values of $\fnl$ even for simulations with $\fnl=0$, such as the ICE-COLA mocks.

\section{DES validation using ICE-COLA mocks}\label{sec:opt_cola}
As mentioned in Section \ref{sec:sims}, the ICE-COLA mocks are designed to match the DES Y3 BAO sample angular mask and redshift distribution $N(z)$. In this Section, we present tests made over the ICE-COLA mocks, assessing their impact on the measurement of the $\fnl$ parameter.

We perform four different tests over the ICE-COLA simulations that we briefly summarise as follows: 
\begin{itemize}
    \item \textbf{Effect of the Integral constraint:} Similarly to Section \ref{sec:test_goliat}, this test double-check the importance of the integral constraint.
    \item \textbf{Best-fit estimator comparison:} This test will tell us how the value of $\fnl$ changes when we consider a different definition for the estimator of the best-fit from the posterior distribution.
    \item \textbf{BAO damping versus Linear theory:} We will show the impact of considering BAO damping in the theoretical modelling by comparing it with the linear power spectrum.
    \item \textbf{Covariance comparison:} For robustness, we consider different covariances and study their impact on the measurement of $\fnl$.
    \item \textbf{Scale configuration:} We compare the effect that different scale cuts and theta binning have when estimating $\fnl$.
\end{itemize}

The fiducial scale configuration for the tests and forecast, along with the optimal $\fnl$ best-fit estimator, are summarised in Table \ref{tab:fiducial_conf}.
The parameters to analyse are presented in detail in the second section of Table \ref{tab:param}. In summary, we consider the linear bias for each redshift bin, the integral constraint as a possible nuisance parameter, and the non-Gaussianity parameter $\fnl$.

\begin{table}
    \centering
    \caption{Fiducial configuration of the ACF for both the \goliat and ICE-COLA mocks.}
    \label{tab:fiducial_conf}
    \begin{tabular}{lcccc} 
	    \hline \hline
	     & $\theta_{\rm min}$ & $\theta_{\rm max}$  & $\Delta \theta$ & $\fnl$ estimator \\ \hline
	     \textbf{\goliat}   & $1.0$ deg. & $20$ deg. & $0.15$ deg. & Max. of marg. posterior\\
	    \hline \\
        \textbf{ICE-COLA} & $1.0$ deg. & $20$ deg. & $0.4$ deg. & Max. of marg. posterior\\
	    \hline
    \end{tabular}
\end{table}
For the analysis, we compare two cases: We perform the MCMC sampling for each mock separately and the mean of the mocks. A summary of the results of this Section is presented in Table \ref{tab:comp_test_COLA}.
The first column presents the mean of the best-fit value of $\fnl$, $\langle\hat{f}_{\rm NL}\rangle$, for the ICE-COLA mocks, obtained from the mean of the best-fit estimator of each mock, $\hat{f}_{\rm NL}$.
The second column presents the overall standard deviation in $\fnl$, obtained from the standard deviation of $\hat{f}_{\rm NL}$ coming from each mock.
The third column is the mean of the $1\sigma$ error obtained from the $\fnl$ posterior of each mock.
The fourth column is the value of $\fnl$ obtained from fitting the theory over the mean of the mocks.
The errors over the mean mocks are from the $68\%$ confidence level of the marginalised posterior distribution.
It is worth remembering that the ICE-COLA mocks have $\fnl=0$ as an initial condition.

The results from this section are presented in Figure \ref{fig:opt_hist}, where for each test, we show the histogram of the best-fit values, $\hat{f}_{\rm NL}$, from each mock. We also show the mean of the histogram, $\langle\hat{f}_{\rm NL}\rangle$, for each test.

After the tests, we forecast the accuracy in the measurement of the local primordial non-Gaussianity parameter $\fnl$ using the angular correlation function with integral constraint over DES Y3 data. We would be able to obtain an error of $\sigma(\fnl)=31$ if the measurement is performed over the DES Y3 BAO sample, as we will see by the end of the section.

\subsection{Effect of the Integral constraint on ICE-COLA mocks}\label{ssec:ic_cola}
Here we show the effect of the IC over the ICE-COLA simulations. We compare the effect of the integral constraint for three different cases:
\begin{itemize}
    \item Without using any integral constraint correction (no IC).    \item Fixing the integral constraint to the value obtained using Eq.(\ref{eq:ic_thetamax}), following the discussion from subsection \ref{ssec:IC_thetamax} (fixed IC).
    \item Considering the integral constraint as a nuisance parameter and marginalising over it (marg IC).
\end{itemize}

\begin{figure}
    \includegraphics[width=\columnwidth]{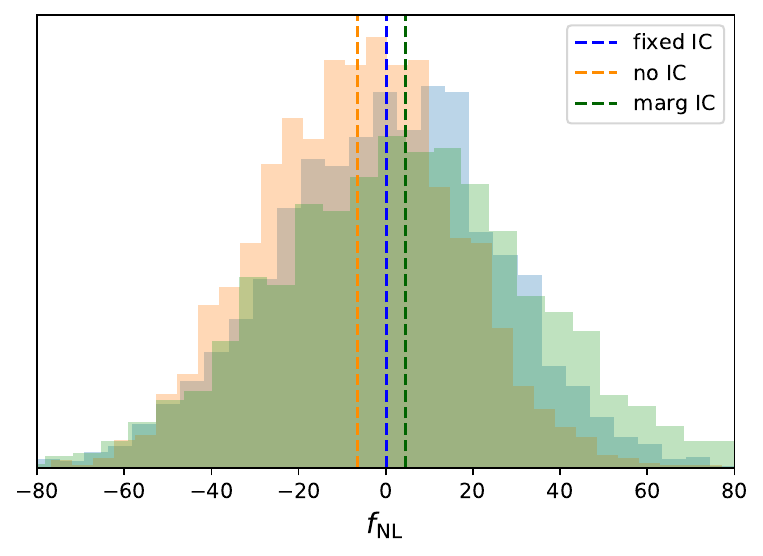}
    \includegraphics[width=\columnwidth]{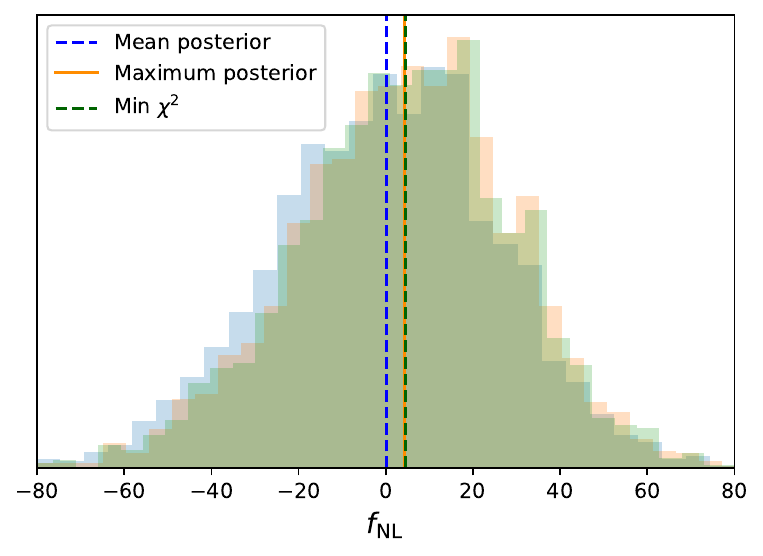}
    \includegraphics[width=\columnwidth]{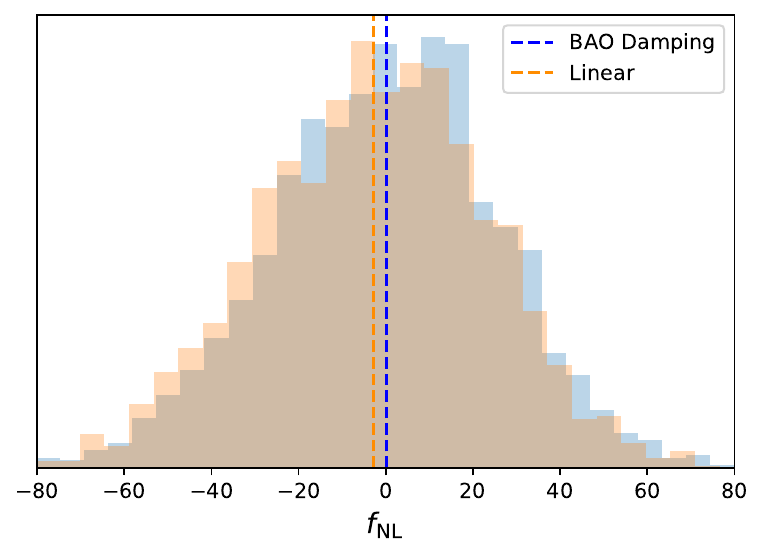}
    \caption{Histograms of the $\fnl$ measurement over the 1952 ICE-COLA mocks comparing the different tests. The vertical dotted lines represent the mean of the histograms. {\bf Top panel}: Effect of the integral constraint. The blue is with fixing the integral constraint as in Eq.(\ref{eq:acf_ic}) (fixed IC), the yellow is without using the integral constraint (no IC), and the green is the integral constraint as a nuisance parameter (marg IC). The vertical dotted lines represent the mean of the histograms. {\bf Middle panel:} Best-fit estimator comparison. The blue is the mean of the posterior as the best-fit, the yellow uses the maximum of the posterior (MAP), and the green uses the minimum of the $\chi^{2}$. {\bf Bottom panel:} Raw linear theory versus BAO damping comparison. The blue includes BAO damping in the template, and the orange uses the linear theory.}
    \label{fig:opt_hist}
\end{figure}
\begin{table}
    \centering
    \caption{Summary of measuring $\fnl$ from the ICE-COLA mocks. The first column is the overall best fit of $\fnl$ obtained as the mean of $\fnl$ from each mock. The second column is the error in $\fnl$ from the standard deviation of every histogram. The third column is the mean of $1\sigma$ error from the $\fnl$ posterior of each mock. The fourth column is the value of $\fnl$ when fitting over the mean of the mocks. The errors are obtained at the $68\%$ confidence level of the posterior. In bold, we highlight the fiducial configuration that will be used for the forecast.}
    \label{tab:comp_test_COLA}
    \begin{tabular}{lcccc} 
	    \hline \hline
	    \textbf{ICE-COLA}\\
	    \hline \hline  &  $\langle \hat{f}_{\rm NL} \rangle$  & std($\hat{f}_{\rm NL}$) &  $\langle \sigma(\hat{f}_{\rm NL}) \rangle$ & mean of mocks\\ \hline \hline
	    NO-IC &  $-7.4$ & $26.6$ & $22.1$ & $-12\pm22$ \\
	    {\bf IC-FIXED} &  $0.1$ & $31$ & $24.8$ & $-4.5\pm24$ \\
	    IC-MARG &  $4.2$ & $35$ & $29$ & $-3\pm27$ \\
	    \hline
	    Mean posterior & $-6.6$ & $30.9$ & -- & -- \\
	    {\bf Max posterior} & $0.1$ & $31$ & -- & -- \\
	    Min $\chi^{2}$ & $0.06$ & $31.1$ & -- & -- \\
	    \hline
	    {\bf Damping} & $0.1$ & $31$ & $24.8$ & $-4.5\pm24$ \\
	    Linear & $2.4$ & $30.5$ & $24.6$ & $-2.2\pm23$ \\
	    \hline
	    {\bf Cosmolike cov.} & $0.1$ & $31$ & $24.8$ & $-4.5\pm24$ \\
	    ICE-COLA cov. & $-0.3$ & $29.6$ & $25.4$ & $-9\pm28$ \\
	    \hline 
	    $w(\theta)[\Delta \theta = 0.1]$ & $-2.2$ & $32.2$ & $25.4$ & $-7.5\pm24$\\
	    $w(\theta)[\Delta \theta = 0.2]$ & $-1.7$ & $32.4$ & $25.5$ & $-6.5\pm24$\\
	    $w(\theta)[\Delta \theta = 0.3]$ & $-0.9$ & $31.8$ & $25.1$ & $-5.5\pm24$\\
	    $\boldsymbol{w(\theta)[\Delta \theta = 0.4]}$ & $0.1$ & $31$ & $24.8$ & $-4.5\pm24$\\
	    \hline
	    $w(\theta)[\theta_{\rm max} = 5]$ & $3.6$ & $35.1$ & $30$ & $-1.7\pm27$\\
	    $w(\theta)[\theta_{\rm max} = 10]$ & $0.6$ & $33.4$ & $26.7$ & $-3.7\pm26$\\
	    $w(\theta)[\theta_{\rm max} = 15]$ & $-0.08$ & $32.4$ & $25.4$ & $-6.2\pm24$\\
	    $\boldsymbol{w(\theta) [\theta_{\rm max} = 20]}$ & $0.1$ & $31$ & $24.8$ & $-4.5\pm24$\\
	    \hline
    \end{tabular}
\end{table}

The results are presented in the top panel of Figure \ref{fig:opt_hist} and summarised in the first part of Table \ref{tab:comp_test_COLA}. From the first column of the table, we notice that not using the integral constraint gives a biased value of $\fnl$, with a deviation of $\Delta\fnl\sim7$ from the fiducial value of the simulation.
We also notice that leaving the integral constraint as a free nuisance parameter gives slightly larger errors for $\fnl$.
Finally, we show that fixing the integral constraint to the value given by Eq.(\ref{eq:ic_thetamax}) gives almost no bias in $\fnl$, recovering the fiducial value of $\fnl=0$ with high accuracy.
Similar to the conclusion from Section \ref{sec:test_goliat}, the integral constraint helps us to avoid biased values of $\fnl$. Although this effect was stronger for non-Gaussian mocks, for the case of $\fnl=0$, we can still notice a difference when measuring $\fnl$.

The mild deviation on $\fnl$ due to not including the integral constraint on ICE-COLA mocks ($\Delta \fnl \sim 7$) opposes the significant bias coming from non-Gaussian mocks ($\Delta\fnl\sim100$). 
Part of this difference is expected to come from a stronger IC effect on smaller mocks (\goliat), but another important effect comes from the IC being stronger mocks with PNG, as we discussed in much detail in Section \ref{sec:ic}.
In order to separate those effects, we now run our fit on a theory-data vector generated for $\fnl=100$ in a DES-like scenario, including the integral constraint and based on the ICE-COLA cosmology.

\begin{figure}
    \includegraphics[width=\columnwidth]{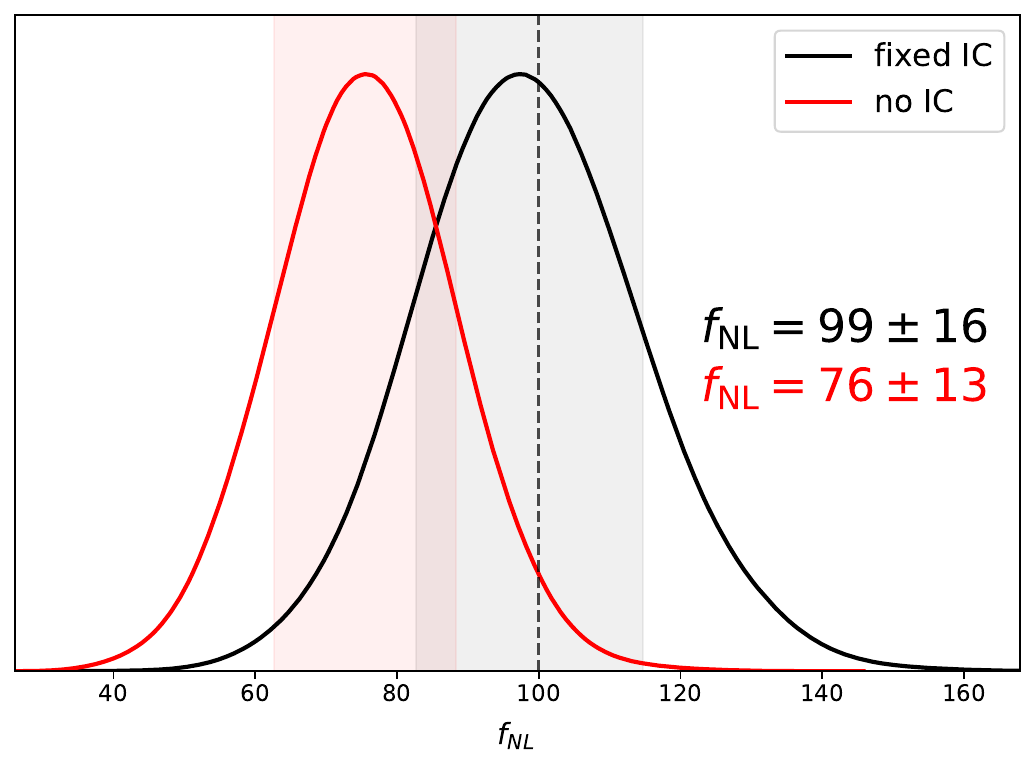}
    \caption{Marginalised $\fnl$ posteriors resulting from fitting our model to a theory-data vector with $\fnl=100$ with integral constraint (IC) and a DES-like setup ($\sim 4100 \rm{deg}^2$, see Table \ref{tab:param}).
    We use both a model with IC (black line) and without IC (red line), finding consistency for the former and a $1.8\sigma$ bias for the latter.
    The shaded areas represent the $\fnl$ marginalized errors at $68\%$ c.l.}
    \label{fig:fnl100td}
\end{figure}

From the posterior distribution of Figure \ref{fig:fnl100td}, we found, as expected, that we recover the fiducial value, $\fnl=99\pm16$, for the case of fixed-IC.
Whereas for the case of ignoring the integral constraint, we found $\fnl=76\pm13$. The deviation of $\Delta \fnl\sim 23$ corresponds to a $1.8\sigma$ bias in the value of $\fnl$ in a non-Gaussian (DES-Y3-like) scenario.
The bias also translates into a mild deviation of $\Delta\chi^{2}\sim 2$ in favour of using the IC in the theoretical model.
Even though the bias on $\fnl$ is not as strong as for the \goliat simulations, we still see a more biased value than the case of $\fnl=0$ simulations.
The same test can be repeated for a theory-data vector with $\fnl=-100$ where we found $\fnl=-95\pm28$ for the fixed-IC case and $\fnl=-80\pm23$ for the no-IC case. In this case, the bias is less significant: $\Delta \fnl\sim 20$, approximately $1\sigma$. 
Hence, even for a large DES-like area, the bias on $\fnl$ when ignoring the IC becomes significant if the data we are fitting contains PNG. 
Similar to the conclusion from Section \ref{sec:test_goliat}, the results highlight the importance of the integral constraint when dealing with primordial non-Gaussianity.

\subsection{Best-fit estimator comparison}
We compare different ways to extract the best-fit estimator $\hat{f}_{\rm NL}$ from the marginalised $\fnl$ posterior distribution, that is, using different central tendency estimators. We show the differences between using the mean of the marginalised posterior, the maximum of the marginalised posterior (MP), or the minimum of the $\chi^{2}$. 

The comparison of the histograms is presented in the second panel of Figure \ref{fig:opt_hist}. The summary of the results from this test is also presented in Table \ref{tab:comp_test_COLA}.
From the table, we can see that we found no considerable differences in using the maximum of the posterior distribution and the minimum of the $\chi^{2}$. Furthermore, we notice an improvement when we use the maximum of the posterior, against the mean of the posterior, as an estimator of the central value for $\fnl$, where we found almost no bias. In the end, the maximum of the posterior was preferred. 

\subsection{Linear theory versus BAO damping}\label{ssec:lin_theory}

As mentioned in the theoretical modelling, we focused on the damping model because of its improvement when fitting the BAO peak.
One open question is whether we need to consider such precision in the template when measuring $\fnl$.

To address the previous question, we compare the $\fnl$ measurement from an ACF with a BAO damping model against using the ACF from a linear power spectrum.
Both ACFs are computed using the fiducial configuration.

We summarise the results in Table \ref{tab:comp_test_COLA}. We found that using a linear power spectrum introduces a small bias compared to including the BAO damping in the power spectrum.

\subsection{Covariance comparison} \label{ssec:cov_comp}

In subsection \ref{ssec:cov}, we mentioned that the default covariance matrix used is Cosmolike since the ICE-COLA presented a spurious correlation between non-adjacent redshift bins. In this subsection, we compare the effect of different covariance in the $\fnl$ measurements. We compare the Cosmolike covariance versus the covariance obtained from the ICE-COLA mocks.

The results are presented in Table \ref{tab:comp_test_COLA}, where we show that the measurement of $\fnl$ is robust against changes in the covariance.

\subsection{Scale configuration}
In this subsection, we discuss the impact of different scale configurations on the measurement of $\fnl$.
We compute the theory and the data vector from each ICE-COLA mock considering a combination of the following scales:
\begin{itemize}
    \item $\theta_{\rm max} = [5, 10, 15, 20]$ deg. 
    \item $\Delta \theta = [0.1, 0.2, 0.3 ,0.4]$ deg.
\end{itemize}

We summarise the extracted information in the last two sections of Table \ref{tab:comp_test_COLA}. For this study, we limited ourselves to a maximum angular separation of 20 degrees because we consider that controlling the LSS systematics up to these scales will already be very challenging. Note that the fiducial maximum angular scale for the BAO measurement was 5 degrees \citepalias{descollaboration2021dark}.

From the results, we can notice two effects. First, the measurements of $\fnl$ seem to be robust against the change in $\Delta \theta$, introducing small changes in both the mean and its error. 
The second effect appears when we go to larger values of $\theta_{\rm max}$, where there is an $\sim 11\%$ improvement in the constraints when going up to $\theta_{\rm max}=20$. This improvement is expected since most of the $\fnl$ effect comes from large scales.

\begin{figure*}
    \includegraphics[width=\textwidth]{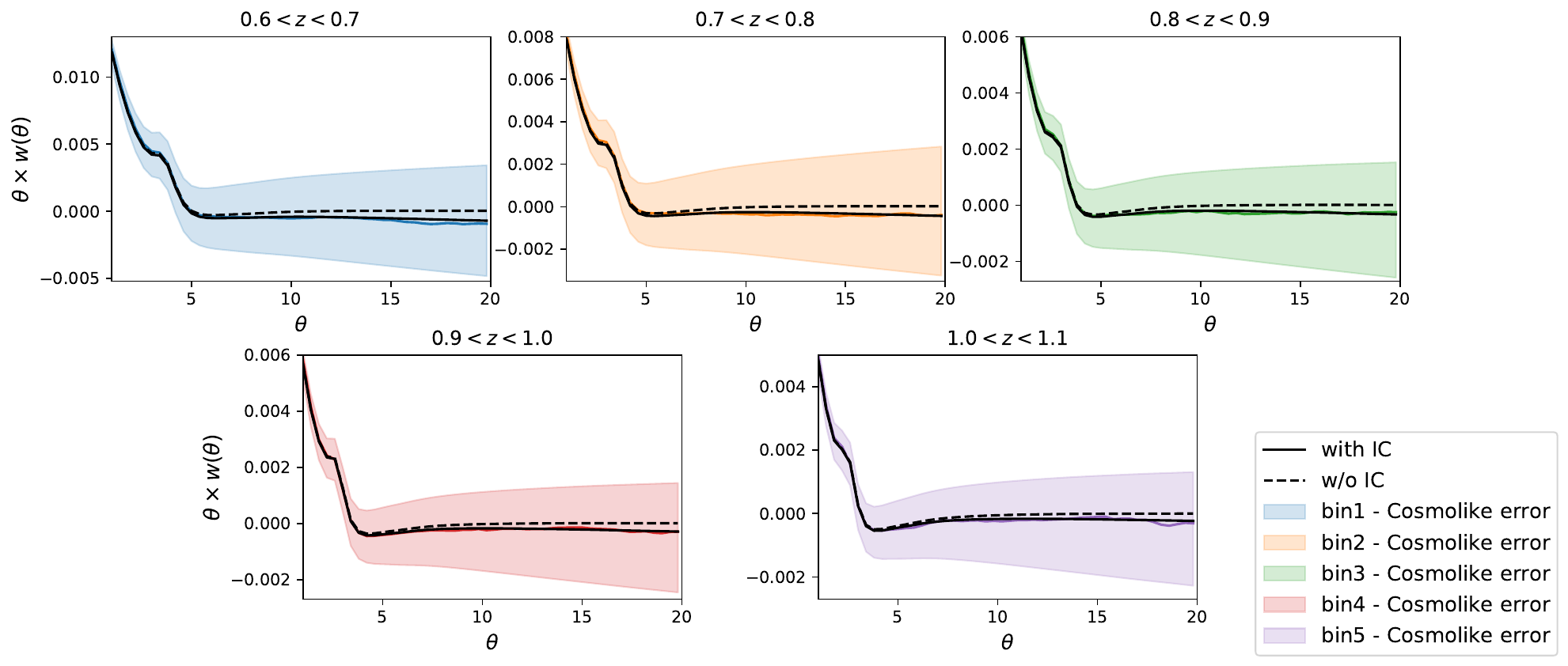}
    \caption{Comparison between the theoretical ACF versus the mean ACF of the ICE-COLA mocks for each redshift bin. The solid black lines are theoretical ACF computed for $\langle\hat{f}_{\rm NL}\rangle$ and $b_g$ obtained using the optimal fiducial configuration and fixing the integral constraint. The black dashed lines are the theoretical ACF without integral constraint. The solid-coloured lines are the mean of the ACF from the mocks. The shaded areas are errors obtained from the COSMOLIKE covariance.}
    \label{fig:acf_cola}
\end{figure*}

From the results of this section, we have three main conclusions: First, we can improve the accuracy of $\fnl$ by using the integral constraint. Not including it is the main source of bias in our measurement, introducing deviations of $\Delta \fnl\sim 7$ to the fiducial value. In the second place, we can improve the precision on $\fnl$ constraints by $\sim 11 \%$ when going to angular scales of $\theta_{\rm max}=20$. Thirdly, our analysis is robust against changes in the type of covariance, the inclusion of BAO damping, and changes in the scale binning, where we found almost no deviations in the precision and accuracy of $\fnl$. 
These conclusions allow us to define the fiducial configuration highlighted in Table \ref{tab:fiducial_conf}.

Using the fiducial configuration, in Figure \ref{fig:acf_cola}, we show the angular correlation function for the best-fit values compared against the mean of the ICE-COLA mocks for each redshift bin with and without the integral constraint, fixed to the value given by Eq.(\ref{eq:ic_thetamax}). From Figure \ref{fig:acf_cola}, we can notice the importance of the integral constraint when comparing the model with the simulations improving its matching, especially at large scales, and therefore, improving the accuracy of $\fnl$.

After the tests from this section, we conclude that a reliable forecast is $\sigma(\fnl)=31$ for the DES Y3 BAO sample after marginalising the linear bias and fixing the other cosmological parameters. The forecast is also done using the fiducial configuration from Table \ref{tab:fiducial_conf}.

\section{Conclusions}
We have presented a methodology to constrain $\fnl$ using the 2-point angular correlation function with scale-dependent bias. Primordial non-Gaussianity modifies the linear bias relation between dark matter overdensities and galaxies by including a scale dependence that depends on the $\fnl$ parameter. The scale dependency is later introduced in the power spectrum and transferred to the ACF. It is worth noticing that there are differences in the effect of the scale-dependent bias; for the power spectrum, the effect is more localised, whereas for the (angular) correlation function, it is more extended over a range of scales.

We remarked on the importance of the integral constraint condition, an observational constraint that appears due to the limited volume observed by surveys and the fact that we estimated the mean number density from them. This condition is essential because of the $\fnl$ effect in the 2-point correlation at large scales and the divergent behaviour of the power spectrum at $k\to0$ (see Eq.\ref{eq:bias_fnl}). We impose the integral constraint condition on our theoretical model and show that it can be corrected by a constant.

We tested the model with the integral constraint correction against the \goliat simulations with non-Gaussian initial conditions. We showed how the integral constraint is a crucial element in avoiding biased $\fnl$ values. 
We showed that ignoring the integral constraint gives very biased PNG constraints, $\fnl = -2.8\pm1.0$ $(\fnl=-10.3\pm1.5)$, whereas we recover the fiducial value $\fnl=100$ $(\fnl=-100)$, within $1\sigma$, when correcting for the integral constraint: $\fnl=97.4\pm3.5$ $(\fnl=-95.2\pm5.4)$.
We confirmed the importance of the integral constraint for simulations with $\fnl=100$ and $\fnl=-100$.

We used the ICE-COLA mocks to validate and test the robustness of the pipeline against different analysis choices when measuring $\fnl$.
We showed that fixing the integral constraint (Eq.(\ref{eq:ic_thetamax})) improves the accuracy in the value of $\fnl$, correcting for a $\Delta \fnl\sim7$ deviation with respect to the fiducial value when not including it.
Furthermore, we showed that going to large angular scales of $\theta_{\rm max}=20$ improves the $\fnl$ precision by $\sim11\%$.
In addition, we showed that not including the BAO damping can introduce a slight bias of $\Delta \fnl\sim 2$.
Also, our results prove to be robust against changes in the choice of covariance matrices and the choice of angular binning.
Using a theory-data vector with $\fnl=100$ ($\fnl=-100$) with IC based on ICE-COLA cosmology, area, and n(z), we also checked the importance of the integral constraint when having the realistic case of a DES-Y3-like survey. 
We found a $\Delta \fnl\sim23$ ($\Delta \fnl\sim15$) deviation when not using the integral constraint in our theoretical modelling.

One of the main conclusions of this paper is that when ignoring the integral constraint in a PNG analysis, we always find a bias on the recovered $\fnl$. This bias is strongest for a small survey and a \textit{true Universe} with PNG (\goliat: $\Delta \fnl\sim 100 \sim \sigma$). For a large survey like DES, we still find a significant bias on $\fnl$ for a \textit{true Universe} with PNG ($\Delta \fnl\sim 20 \sim 1-2 \sigma$, for $\fnl^{\rm true}=100$). Whereas the bias on $\fnl$ is mild for a large survey ($\sim 4100$deg$^2$) and a Gaussian \textit{true Universe} ($\Delta \fnl\sim 7 \sim 0.3 \sigma$).

We expect our analysis to be the first step into constraining $\fnl$ with the Dark Energy Survey photometric data, where we forecast a measurement of $\fnl$ within $\sigma(\fnl)=31$ when measured against the DES Y3 BAO sample.
This prospect is comparable with the current constraints coming from spectroscopic surveys, being $\sigma(\fnl)\sim 21$ \citep{mueller2021clustering} the latest one to date.

Future plans include mitigation of LSS systematics following up on \citet{rosell2021dark,rodriguezmonroy2021dark} with a particular focus on very large scales \citep[see, e.g.][]{2021rezaie}, which is crucial as systematic errors due to survey properties can lead to spurious PNG signal \citep{Ross2013, mueller2021clustering}. Given this, we plan to conduct a full battery of robustness tests while blinded to the $\fnl$ value, following the standard DES policy \citepalias[e.g.]{descollaboration2021dark}. Additionally, performing PNG analysis can also be understood as a strong validation exercise of the galaxy clustering systematics, given the sensitivity of this probe to them. Note also that future photometric surveys are expected to break the barrier of $\sigma(\fnl)=1$ \citep{PhysRevD.95.123513}, key to the inflationary models, and this work is a necessary step toward that goal.

\section*{Acknowledgements}

The work of WR is funded by a fellowship from “La Caixa” Foundation (ID 100010434) with fellowship code LCF/BQ/DI18/11660034 and the European Union Horizon 2020 research and innovation programme under the Marie Skłodowska-Curie grant agreement No. 713673. 
SA is currently supported by the Spanish Agencia Estatal de Investigacion through the grant “IFT Centro de Excelencia Severo Ochoa by CEX2020-001007-S", he was also supported by the MICUES project, funded by the EU H2020 Marie Skłodowska-Curie Actions grant agreement no. 713366 (InterTalentum Fellowship UAM). WR, SA and JGB acknowledge support from the Research Project PGC2018-094773-B-C32 and the Centro de Excelencia Severo Ochoa Program CEX2020-001007-S.
WR, SA and JGB acknowledge the use of the Hydra cluster at IFT.

Funding for the DES Projects has been provided by the U.S. Department of Energy, the U.S. National Science Foundation, the Ministry of Science and Education of Spain, 
the Science and Technology Facilities Council of the United Kingdom, the Higher Education Funding Council for England, the National Center for Supercomputing 
Applications at the University of Illinois at Urbana-Champaign, the Kavli Institute of Cosmological Physics at the University of Chicago, 
the Center for Cosmology and Astro-Particle Physics at the Ohio State University,
the Mitchell Institute for Fundamental Physics and Astronomy at Texas A\&M University, Financiadora de Estudos e Projetos, 
Funda{\c c}{\~a}o Carlos Chagas Filho de Amparo {\`a} Pesquisa do Estado do Rio de Janeiro, Conselho Nacional de Desenvolvimento Cient{\'i}fico e Tecnol{\'o}gico and 
the Minist{\'e}rio da Ci{\^e}ncia, Tecnologia e Inova{\c c}{\~a}o, the Deutsche Forschungsgemeinschaft and the Collaborating Institutions in the Dark Energy Survey. 

The Collaborating Institutions are Argonne National Laboratory, the University of California at Santa Cruz, the University of Cambridge, Centro de Investigaciones Energ{\'e}ticas, 
Medioambientales y Tecnol{\'o}gicas-Madrid, the University of Chicago, University College London, the DES-Brazil Consortium, the University of Edinburgh, 
the Eidgen{\"o}ssische Technische Hochschule (ETH) Z{\"u}rich, 
Fermi National Accelerator Laboratory, the University of Illinois at Urbana-Champaign, the Institut de Ci{\`e}ncies de l'Espai (IEEC/CSIC), 
the Institut de F{\'i}sica d'Altes Energies, Lawrence Berkeley National Laboratory, the Ludwig-Maximilians Universit{\"a}t M{\"u}nchen and the associated Excellence Cluster Universe, 
the University of Michigan, NSF's NOIRLab, the University of Nottingham, The Ohio State University, the University of Pennsylvania, the University of Portsmouth, 
SLAC National Accelerator Laboratory, Stanford University, the University of Sussex, Texas A\&M University, and the OzDES Membership Consortium.

Based in part on observations at Cerro Tololo Inter-American Observatory at NSF's NOIRLab (NOIRLab Prop. ID 2012B-0001; PI: J. Frieman), which is managed by the Association of Universities for Research in Astronomy (AURA) under a cooperative agreement with the National Science Foundation.

The DES data management system is supported by the National Science Foundation under Grant Numbers AST-1138766 and AST-1536171.
The DES participants from Spanish institutions are partially supported by MICINN under grants ESP2017-89838, PGC2018-094773, PGC2018-102021, SEV-2016-0588, SEV-2016-0597, and MDM-2015-0509, some of which include ERDF funds from the European Union. IFAE is partially funded by the CERCA program of the Generalitat de Catalunya.
Research leading to these results has received funding from the European Research
Council under the European Union's Seventh Framework Program (FP7/2007-2013) including ERC grant agreements 240672, 291329, and 306478.
We  acknowledge support from the Brazilian Instituto Nacional de Ci\^encia
e Tecnologia (INCT) do e-Universo (CNPq grant 465376/2014-2).

This manuscript has been authored by Fermi Research Alliance, LLC under Contract No. DE-AC02-07CH11359 with the U.S. Department of Energy, Office of Science, Office of High Energy Physics.

\section*{Data availability}

The data underlying this article will be shared on reasonable request to the corresponding author.



\bibliographystyle{mnras}
\bibliography{png_acf_des} 





\appendix

\section{The analytic correlation function, the IC and $\fnl$}\label{ap:acf_theory}

In order to gain some insight about divergent behavior mentioned in subsection \ref{ssec:theo_ic} due to the theoretical integral constraint with $\fnl$, in this appendix, we compute the explicit dependence on the large scales of the integral constraint condition for the 2-point correlation function. 

Let us start by considering the primordial power spectrum $P_{\Phi}(k) = A\,k^{n_{s}}$, which can be used to define the linear matter power spectrum by considering a simplified transfer function \citep{peacock_1998},

\begin{equation}
    P_{m}(k) = P_{\Phi}(k) \,T^2(k) =
    \frac{A\,k^{n_{s}}}{(1+k^2/\keq^2)^2},
\end{equation}
where $\keq$ is the wavenumber at matter-radiation equality. 
As previously seen in Section \ref{sec:theory}, from the matter power spectrum, we can compute the multipole expansion of the two-point correlation function using Eq.(\ref{eq:2PCF}). 

For simplicity, we focus on the monopole. It is possible to compute the 2PCF for $n=1$ and $\keq=1$ as
\begin{eqnarray}
    \xi_0(r) &=& \frac{1}{4\pi^2 r}\, 
    \left(g(r) + r\,g'(r)\right)\,,\\ \label{eq:xi_0_a}
    g(r) &=& \cosh\,r \,\shi\,r - \sinh\,r \,\CHI\,r\,,
\end{eqnarray}
where $\shi()$ and $\CHI()$ are the Sinh- and Cosh-Integral functions. This shows that the 2PCF can be computed analytically for this power spectrum. 

The next step is to show analytically how the 2PCF changes if we include a scale-dependent bias and use the simplified matter power spectrum. Let us start by recalling the expression of the power spectrum with scale-dependent bias,
\begin{eqnarray}
    P_{g}(k) &=& b(k)^2 P_{m}(k), \\[1mm]
    b(k) &=& b_{g} + \Delta b(k, z).
\end{eqnarray}
We find terms that are independent, linear, and quadratic in $\fnl$. This implies that the computation of the 2PCF involves three integrals over the wavenumbers.
The term \textit{independent of $\fnl$} just gives something proportional to $b_{g}^2\,\xi_0(r)$.

The \textit{linear term in $\fnl$} is more interesting. This component of 2PCF is proportional to, 
\begin{equation}\label{eq:ap_lin}
    \int_0^\infty  \text{d}k k^2 \frac{P_{m}(k)}{k^2\,T(k)} j_0(kr) = \frac{g(r)}{r},
\end{equation}
which is finite for large values of $r$.

The \textit{quadratic term in $\fnl$} logarithmically diverges as $\kmin\to0$, this can be seen as follows,
\begin{eqnarray}\label{eq:ap_quad}
    & {\displaystyle 
    \int_{\kmin}^\infty k^2 \frac{P_{m}(k)}{k^4\,T(k)^2} \,j_0(k\,r) \text{d}k =
    j_0(\kmin r) - {\rm ci}(\kmin r)} \\
    & {\displaystyle\to 1-\gamma-\ln(\kmin r) + \frac{1}{12}\kmin^2r^2 }, \label{xi_quad}
\end{eqnarray}
where ci is the Cosine Integral function.

Now that we have computed the 2PCF for the simplified power spectrum with scale-dependent bias, we can analyze how the theoretical integral constraint condition, given by Eq.(\ref{eq:pk_0}), behaves at large scales.

From the previous computation can be seen that the integral constraint condition explicitly vanishes for the term independent of $\fnl$,
\begin{equation}
    \int_0^\infty \text{d}r r^2\xi_0(r) = 0.
\end{equation}

The linear term, given by Eq.(\ref{eq:ap_lin}) is linearly divergent for a given large scale $r_{max}$. This can be seen as follows,
\begin{equation}
    \int_0^{\rmax} \text{d}r r^2 \frac{g(r)}{r} =
    \rmax(1+g'(\rmax)) - g(\rmax)
    \to \rmax \,, \label{eq:div_xi}
\end{equation}
This implies that there will be a linear term in $\fnl$ proportional to $\fnl\rmax$.

Now if we compute the integral constraint for the quadratic term in $\fnl$, given by Eq.(\ref{eq:ap_quad}), we find that is proportional to $\fnl^2\keq^3/\kmin^3$.

Therefore, from this calculation, we conclude that the integral constraint has a term linear in $\fnl$, which diverges with $\rmax$, and a quadratic term in $\fnl^2$ which is proportional to $\kmin^{-3}$.
This implies that, even for infinite volume surveys, we need to correct the two-point correlation function with PNG with the integral constraint, because it can bias the $\fnl$ results.

\bsp	
\label{lastpage}
\end{document}